\documentclass[aps,prb,twocolumn,showpacs,10pt,superscriptaddress]{revtex4-1}
\usepackage{graphicx}
\usepackage{amssymb}
\usepackage{amsmath}
\usepackage{comment}
\usepackage{color}
\pdfoutput=1

\begin{document}

\def\pc{\frac{2\pi}{\Phi_0}}

\def\e{\varepsilon}
\def\f{\varphi}
\def\p{\partial}
\def\ba{\mathbf{a}}
\def\bA{\mathbf{A}}
\def\bb{\mathbf{b}}
\def\bB{\mathbf{B}}
\def\bD{\mathbf{D}}
\def\bd{\mathbf{d}}
\def\be{\mathbf{e}}
\def\bE{\mathbf{E}}
\def\bH{\mathbf{H}}
\def\bj{\mathbf{j}}
\def\bk{\mathbf{k}}
\def\bK{\mathbf{K}}
\def\bM{\mathbf{M}}
\def\bm{\mathbf{m}}
\def\bn{\mathbf{n}}
\def\bq{\mathbf{q}}
\def\bp{\mathbf{p}}
\def\bP{\mathbf{P}}
\def\br{\mathbf{r}}
\def\bR{\mathbf{R}}
\def\bS{\mathbf{S}}
\def\bu{\mathbf{u}}
\def\bv{\mathbf{v}}
\def\bV{\mathbf{V}}
\def\bw{\mathbf{w}}
\def\bx{\mathbf{x}}
\def\by{\mathbf{y}}
\def\bz{\mathbf{z}}
\def\bG{\mathbf{G}}
\def\bW{\mathbf{W}}
\def\Bn{\boldsymbol{\nabla}}
\def\Bo{\boldsymbol{\omega}}
\def\Br{\boldsymbol{\rho}}
\def\Bs{\boldsymbol{\hat{\sigma}}}
\def\bh{{\beta\hbar}}
\def\mA{\mathcal{A}}
\def\mB{\mathcal{B}}
\def\mD{\mathcal{D}}
\def\mF{\mathcal{F}}
\def\mG{\mathcal{G}}
\def\mH{\mathcal{H}}
\def\mI{\mathcal{I}}
\def\mL{\mathcal{L}}
\def\mO{\mathcal{O}}
\def\mP{\mathcal{P}}
\def\mT{\mathcal{T}}
\def\mU{\mathcal{U}}
\def\mZ{\mathcal{Z}}
\def\fr{\mathfrak{r}}
\def\ft{\mathfrak{t}}
\newcommand{\rf}[1]{(\ref{#1})}
\newcommand{\al}[1]{\begin{aligned}#1\end{aligned}}
\newcommand{\ar}[2]{\begin{array}{#1}#2\end{array}}
\newcommand{\eq}[1]{\begin{equation}#1\end{equation}}
\newcommand{\bra}[1]{\langle{#1}|}
\newcommand{\ket}[1]{|{#1}\rangle}
\newcommand{\av}[1]{\langle{#1}\rangle}
\newcommand{\AV}[1]{\left\langle{#1}\right\rangle}
\newcommand{\aav}[1]{\langle\langle{#1}\rangle\rangle}
\newcommand{\braket}[2]{\langle{#1}|{#2}\rangle}
\newcommand{\ff}[4]{\parbox{#1mm}{\begin{center}\begin{fmfgraph*}(#2,#3)#4\end{fmfgraph*}\end{center}}}

\def\mr{m_{\perp}}
\def\ml{m_{\parallel}}
\def\hr{H_{\perp}}
\def\hl{H_{\parallel}}

\def\mb{(\mu+\alpha\nu)}
\def\nb{(\nu-\alpha\mu)}
\def\lb{(\lambda+\alpha\kappa)}
\def\kb{(\kappa-\alpha\lambda)}
\def\mn{\left|\bm\times\bz\right|}
\def\etap{\frac{2\pi}{\Phi_0}}
\def\ab{\bar{\alpha}}

\title{Nonequilibrium spin current through interacting quantum dots}

\author{Silas Hoffman}
\affiliation{Department of Physics and Astronomy, University of California, Los Angeles, California 90095, USA}
\affiliation{Department of Physics, University of Basel, Klingelbergstrasse 82, CH-4056 Basel, Switzerland}
\author{Yaroslav Tserkovnyak}
\affiliation{Department of Physics and Astronomy, University of California, Los Angeles, California 90095, USA}

\begin{abstract}
We develop a theory for charge and spin current between two canted magnetic leads flowing through a quantum dot with an arbitrary local interaction. For a noncollinear magnetic configuration, we calculate equilibrium and nonequilibrium current biased by voltage or temperature difference or pumped by magnetic dynamics. We are able to explicitly separate the equilibrium and nonequilibrium contributions to the current, both of which can be written in terms of the full retarded Green's function on the dot. Taking the specific example of a single-level quantum dot with a large on-site Coulomb interaction, we calculate the total spin current near the Kondo regime, which we find to be generally enhanced in magnitude as compared to the noninteracting case.
\end{abstract}



\maketitle

\section{Introduction}

Electron transport is an important tool for probing and manipulating magnetic heterostructures. In a ferromagnet$\mid$insulator$\mid$ferromagnet (F$\mid$I$\mid$F) junction, the charge conductance can be used to measure the magnetic misalignment in the ferromagnetic layers using tunnel magnetoresistance;\cite{jullierePLA75} equilibrium spin current between the two magnets characterizes their interlayer exchange coupling;\cite{slonczewskiPRB89} and nonequilibrium spin current induces a spin-transfer torque on a ferromagnetic layer noncollinear with the direction of spin polarization of the current.\cite{,slonczewskiPRB89,slonczewskiJMMM96,*bergerPRB96}

When impurities are present in the tunneling layer of a magnetic tunnel junction, both equilibrium\cite{jansenAPL99,*jansenPRB00,*velevAPL07,*miaoPRL08,tsymbalPRL03} and nonequilibrium\cite{waintalPRL03,*birolPRB09,manchonJPCM07} spin current can be strongly modified. Specifically, there is an enhancement of spin current when the voltage is tuned to the energy level of the impurity.\cite{tsymbalPRL03,manchonJPCM07} Magnetic impurities strongly tunnel-coupled to metallic leads are known to generate a peak in the density of states of the impurity, called the Kondo resonance, at the Fermi level of the leads,\cite{hewsonBK97} thereby inducing a peak in charge conductance at zero bias. If the leads are magnetic, the Kondo resonance is perturbed by the exchange interaction in the leads,\cite{martinekPRL03b,*choiPRL04,*simonPRB07} splitting the zero-bias anomaly in the differential conductance in both collinear\cite{lopezPRL03,*martinekPRL03,*pasupathySCI04,*utsumiPRB05,*barnasJPCM08} and noncollinear\cite{swirkowiczPRB06} orientations of the leads. It was, furthermore, theoretically shown that spin accumulation associated with a spin-dependent bias also splits the Kondo resonance, which manifests in the charge and spin current.\cite{baoEPL08,*swirkowiczJMMM09,*limPRB13}

In this paper, we consider equilibrium as well as nonequilibrium charge and spin currents through quantum-dot islands or impurities connecting \textit{noncollinear} ferromagnetic leads. When the magnetic leads are aligned along the same axis, the equilibrium spin current vanishes (neglecting spin-orbit interactions). However, for a noncollinear alignment, the spin current is generally finite even in equilibrium. Out of equilibrium, exchange interactions lead to additional voltage-induced torques on the leads. Furthermore, a dynamic reorientation of magnetic leads contributes to an effective spin-dependent chemical potential in the leads in addition to spin accumulation.\cite{brataasPRB02,*tserkovPRB08} We calculate the full out-of-equilibrium spin and charge currents between two canted magnetic leads separated by a quantum dot with a spin-independent on-site interaction. Although related expressions have been derived for charge\cite{meirPRL92} and spin currents,\cite{braunSM05,zhangJPCM09} our generalization accommodates spin-dependent bias and microwave precession of magnetic contacts, in addition to voltage and temperature difference applied to the leads.

We specialize our formalism to two extreme cases: a dot with (1) no on-site interaction and (2) large Coulomb repulsion. As compared with the noninteracting case, we find a large enhancement in equilibrium and nonequilibrium spin currents for the interacting dot. In addition to the peaks in the differential charge conductance as a function of voltage bias at the standard spin-split Kondo resonances, we find subsidiary peaks at twice the corresponding bias due to a partial restoration of the original Kondo resonance; the spin conductance analogously shows features at these biases. Charge and spin pumping driven by magnetic dynamics are, likewise, strongly modified by the presence of on-site Coulombic interaction which we discuss in detail below.

The paper is organized as follows: In the next section, we describe our formalism and derive general expressions used to calculate charge and spin current. Section~\ref{green} fleshes out the retarded Green's functions for the special cases of the free and Kondo dots, recapping standard results from the literature. In Sec.~\ref{equilibrium}, these Green's functions are used to calculate the polarization of the dot and spin current in equilibrium, as a function of relative angle between the magnetic leads, comparing the interacting and noninteracting dots. The nonequilibrium nonlinear charge and spin differential conductances are calculated in Secs.~\ref{voltage} and \ref{spin}, as a function of applied voltage and frequency of the microwave precession, respectively. We summarize our results and suggest additional systems that may be amenable to our formalism in Sec.~\ref{summary}.
 
\section{General Spin-Current Expression}

\begin{figure}[pt]
\includegraphics[width=0.9\linewidth,clip=]{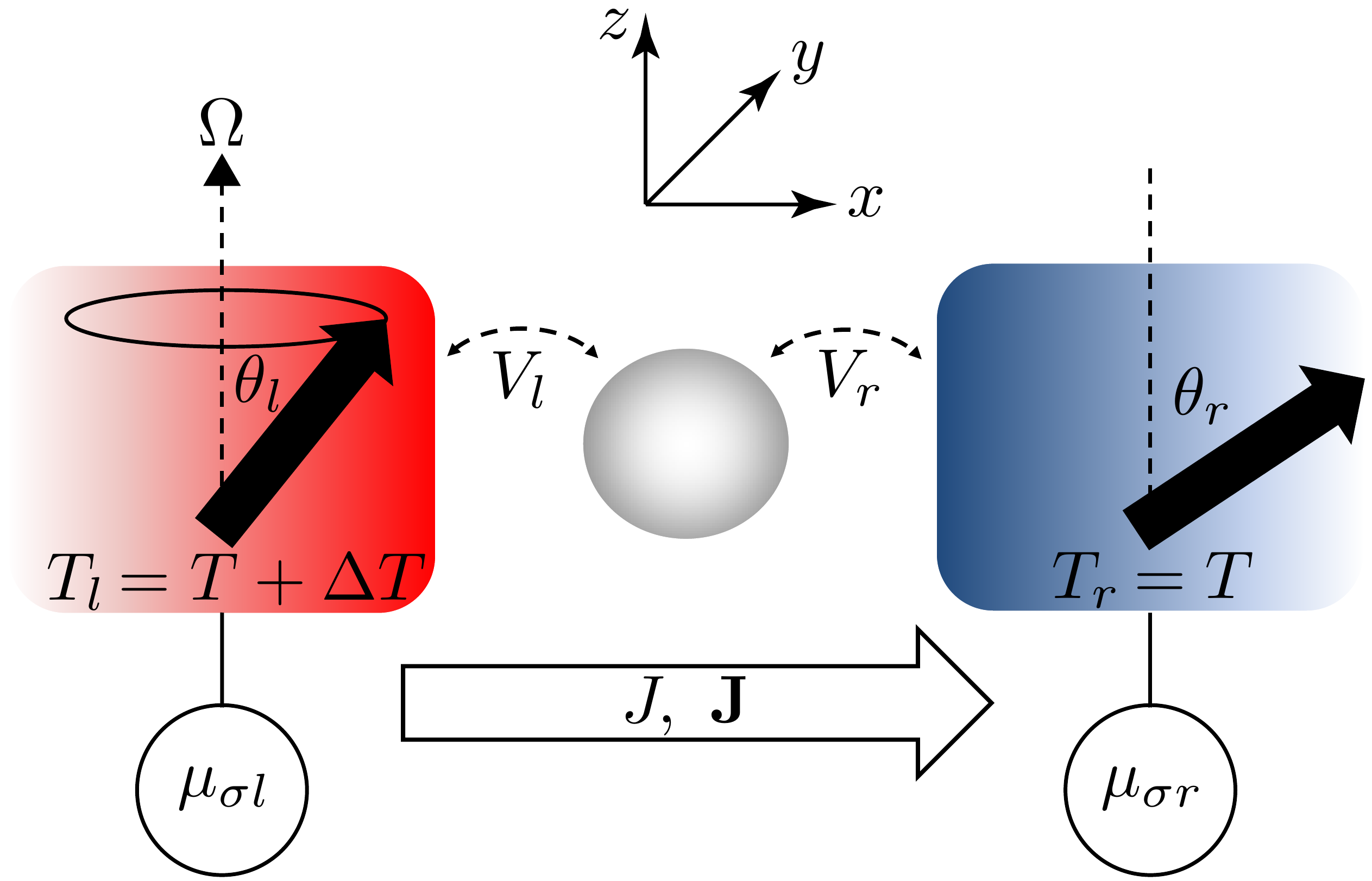}
\caption{Schematic of our model, in which two noncollinear magnetic leads are held at separate temperatures ($T_l$ and $T_r$) and spin-dependent chemical potentials ($\mu_{\sigma l}$ and $\mu_{\sigma r}$, with $\sigma={\uparrow,\downarrow}$). When in contact with a quantum dot, tunneling (with amplitudes $V_l$ and $V_r$ from the left and right leads, respectively) facilitates charge current, $J$, and spin current, $\textbf{J}$, flowing through the dot. The left (right) lead is canted clockwise in the $xz$ plane at an angle $\theta_l$ ($\theta_r$) with respect to the $z$ axis. In Sec.~\ref{spin}, we, additionally, allow the magnetization in the left lead to precess with frequency $\Omega$.}
\label{device}
\end{figure}

We consider a Hamiltonian of the form
\begin{equation}
H=H_{L}+H_{D}+H_{T}\,,
\end{equation}
which describes Fermi-liquid leads ($H_L$) coupled to a quantum dot ($H_D$) by tunneling ($H_T$), defined as follows:
\begin{equation}\begin{aligned}
H_{L}&=\sum_{\sigma k\gamma}\varepsilon_{\sigma k\gamma} c_{\sigma k\gamma}^{\dagger}c_{\sigma k\gamma} \,,\\H_{T}&=\sum_{i \sigma k\gamma} V_{i\sigma k\gamma}d_i^\dagger c_{\sigma k\gamma}+{\rm H.c.}\,.
\end{aligned}\end{equation}
The summation in the lead part of the Hamiltonian is over spin ($\sigma={\uparrow},{\downarrow}$ quantized along the respective magnetic orientation), momentum ($k$), and left ($\gamma=l$) and right ($\gamma=r$) leads.  $\varepsilon_{\sigma k\gamma}$ is the energy of the corresponding electron that is created (annihilated) by $c_{\sigma k\gamma}^{\dagger}$ ($c_{\sigma k\gamma}$). See Fig.~\ref{device} for a schematic of our setup. The dot Hamiltonian is a Hermitian function of dot creation and annihilation operators $d_i^\dagger$ and $d_j$, respectively, that furnish a complete basis for the local Fock space:
\eq{
H_D=\sum_{i}\epsilon_i d^\dagger_i d_i + \mathcal{U}(d^\dagger_i, d_j)\,,
}
where $\mathcal{U}(d^\dagger_i, d_j)$ is a spin-isotropic electron-electron interaction on the dot. The indices $i$ and $j$ run over the relevant quantum numbers. As the simplest nontrivial example, we consider a single spin-1/2 on-site energy level, so that $i,j={\uparrow},{\downarrow}$ for spin along the $z$ axis. Note that the dot can in general be Zeeman split by an external magnetic field or by the proximity-induced exchange field from the magnetic leads. Although we have chosen a single level on the dot connected to two leads, our formalism can be readily generalized to an arbitrary number of levels and leads.

We take two noncollinear magnetic leads, whose orientations define the $xz$ plane, as sketched in Fig.~\ref{device}, which mix the majority and minority electron bands via tunneling. The matrix defining the hopping in the tunneling Hamiltonian is
\eq{\hat{V}_{k\gamma}\equiv
\left(
\begin{array}{cc}
V_{\uparrow k\gamma}\cos\frac{\theta_\gamma}{2} & -V_{\downarrow k\gamma}\sin\frac{\theta_\gamma}{2} \\ 
V_{\uparrow k\gamma}\sin\frac{\theta_\gamma}{2} & V_{\downarrow k\gamma}\cos\frac{\theta_\gamma}{2}
\end{array}
\right)\,,\label{hV}}
where $\theta_\gamma$ is the angle of lead $\gamma$ with respect to the $z$ axis. $V_{\uparrow k\gamma}$ ($V_{\downarrow k\gamma}$) is the matrix element for a spin-preserving spin-up (down) electron tunneling from lead $\gamma$ to the dot.

\begin{widetext}
Using Keldysh formalism,\cite{meirPRL92} we calculate charge and spin currents out of lead $\gamma$ under an arbitrary bias:
\begin{equation}
\begin{aligned}
J_\gamma&=\frac{2e}{h}\mbox{Re}\int d\omega \mbox{Tr}\left[\hat W^{<}_\gamma(\omega)\hat G^{R}(\omega)-\hat W^{R}_\gamma(\omega) \hat G^<(\omega)\right]\,,\\
\textbf{J}_\gamma&=\mbox{Re}\int \frac{ d\omega}{2\pi} \mbox{Tr}\left[\hat W^{<}_\gamma(\omega)\hat{\boldsymbol{\sigma}} \hat G^{R}(\omega)-\hat W^{R}_\gamma(\omega) \hat{\boldsymbol{\sigma}} \left(\hat G^<(\omega)-\frac{i}{\pi}\mP\int d\omega' \frac{\hat{G}^<(\omega')}{\omega'-\omega}\right)\right]\,,
\label{neq_curr}
\end{aligned}
\end{equation}
where $e<0$ is the electron charge. $\hat{G}^R$ and $\hat{G}^<$ are respectively the Fourier-transformed retarded and lesser Green's functions on the dot and $\hat{G}^A=(\hat{G}^R)^\dagger$ is the advanced Green's function. $\mP$ denotes the Cauchy principal value of the integral and $\hat{\boldsymbol{\sigma}}$ is a vector of Pauli matrices. We have defined
\begin{align}
\hat{W}^R_{\gamma}&=-\frac{i}{2}\left(
\begin{array}{cc}
\Gamma_{\uparrow\gamma}\cos^2\frac{\theta_\gamma}{2}+\Gamma_{\downarrow\gamma}\sin^2\frac{\theta_\gamma}{2} & \frac{1}{2}(\Gamma_{\uparrow\gamma}-\Gamma_{\downarrow\gamma})\sin\theta_\gamma \\ 
\frac{1}{2}(\Gamma_{\uparrow\gamma}-\Gamma_{\downarrow\gamma})\sin\theta_\gamma & \Gamma_{\downarrow\gamma}\cos^2\frac{\theta_\gamma}{2}+\Gamma_{\uparrow\gamma}\sin^2\frac{\theta_\gamma}{2}
\end{array}
\right)
\end{align}
and
\begin{align}
\hat{W}^<_{\gamma}&=i\left(
\begin{array}{cc}
f_{\uparrow\gamma}(\omega)\Gamma_{\uparrow\gamma}\cos^2\frac{\theta_\gamma}{2}+f_{\downarrow\gamma}(\omega)\Gamma_{\downarrow\gamma}\sin^2\frac{\theta_\gamma}{2} & \frac{1}{2}[f_{\uparrow\gamma}(\omega)\Gamma_{\uparrow\gamma}-f_{\downarrow\gamma}(\omega)\Gamma_{\downarrow\gamma}]\sin\theta_\gamma \\ 
\frac{1}{2}\left[f_{\uparrow\gamma}(\omega)\Gamma_{\uparrow\gamma}-f_{\downarrow\gamma}(\omega)\Gamma_{\downarrow\gamma}\right]\sin\theta_\gamma & f_{\downarrow\gamma}(\omega)\Gamma_{\downarrow\gamma}\cos^2\frac{\theta_\gamma}{2}+f_{\uparrow\gamma}(\omega)\Gamma_{\uparrow\gamma}\sin^2\frac{\theta_\gamma}{2}
\end{array}
\right)\,.
\end{align}
$\Gamma_{\sigma\gamma}(\omega)=2\pi\sum_k \left|V_{\sigma k\gamma}\right|^2 \delta(\omega-\varepsilon_{\sigma k\gamma})$ and $f_{\sigma\gamma}(\omega)$ is the Fermi-Dirac distribution function of lead $\gamma$ in the majority ($\sigma={\uparrow}$) or minority ($\sigma={\downarrow}$) band. $\Gamma_{\sigma\gamma}$ will be taken to be energy independent within the electron bandwidth and zero otherwise. The details of our derivation are straightforward and can be found in appendix \ref{dgc}.

The following substitutions will be convenient (at the Fermi level): $\Gamma_{\uparrow(\downarrow)l}=\Gamma(1+\alpha)(1\pm p_l)$ and $\Gamma_{\uparrow(\downarrow)r}=\Gamma(1-\alpha)(1\pm p_r)$, where the polarization $p_\gamma$ of lead $\gamma$ may take values from zero (normal metal) to one (halfmetal).  We may vary the weight of hopping from left and right leads to the dot by changing $\alpha$ from $-1$ (coupled only to the right lead) to $1$ (coupled only to the left lead), while $\alpha=0$ is the left-right symmetric case. This allows us to rewrite 
\eq{\hat{W}^{R}_\gamma=-\frac{i(1\pm\alpha)\Gamma}{2}\left(
\begin{array}{cc}
1+p_\gamma \cos\theta_\gamma & p_\gamma\sin\theta_\gamma \\ 
p_\gamma\sin\theta_\gamma & 1-p_\gamma \cos\theta_\gamma 
\end{array}
\right)\,,}
with $\pm$ corresponding to the left (right) lead.

A serious challenge in the calculation of the nonequilibrium current is determining the on-site lesser Green's function. One may use Dyson's equation and Keldysh formalism to show that\cite{jauhoPRB94} $\hat G^<=\hat G^R\hat\Sigma^< \hat G^A$, out of equilibrium, where $\hat\Sigma^<$ stands for the dot's lesser self-energy due to tunneling and interactions. However, because $\hat\Sigma^<$ is a functional of $\hat G^<$, it is in general difficult to obtain an analytic expression for $\hat G^<$. Consequently, we use a generalization\cite{swirkowiczPRB06} to Ng's ansatz,\cite{ngPRL96} taking into account matrix-valued Green's functions: $\hat\Sigma^< = -(\hat W^<_l + \hat W^<_r)(\hat W^R_l + \hat W^R_r)^{-1}[(\hat G^R)^{-1}-(\hat G^A)^{-1}]/2$. This ansatz recovers the exact result for	$\hat{G}^<$ when either no bias is applied or there is no interaction on the dot. For a better ansatz one could anti-Hermitize $\hat\Sigma^<$, which yields an equivalent expression for charge current, but we use the simplified form for the numerics. In the special case when the bias is spin independent, $\hat{W}_\gamma^<=-2f_\gamma \hat{W}_\gamma^R$ and, using Ng's ansatz, we can separate the equilibrium (first line) and nonequilibrium (second line) contributions to the spin current:
\begin{align}
\label{s_curr}
\textbf{J}_\gamma = -\textnormal{Re}\int\frac{ d\omega}{2\pi} \textnormal{Tr}\Bigg[&\left\{f_\gamma(\omega)\hat{W}_\gamma^{R}\hat{\boldsymbol{\sigma}} \left[\hat G^R(\omega) + \hat G^A(\omega)\right]-\hat W^{R}_\gamma \hat{\boldsymbol{\sigma}}\frac{i}{\pi}\mP\int d\omega' \frac{\hat{G}^<(\omega')}{\omega'-\omega}\right\}\Bigg] \nonumber\\
&+\left[f_{\bar\gamma}(\omega)-f_\gamma(\omega)\right]\left\{\hat{W}^{R}_\gamma \hat{\boldsymbol{\sigma}}\hat{G}^R(\omega)\hat W^R_{\bar\gamma}\left(\hat W^R_l + \hat W^R_r\right)^{-1}\left[\hat G^R(\omega)^{-1}-\hat G^A(\omega)^{-1}\right]\hat{G}^A(\omega)\right\}\,,
\end{align}
where lead $\bar\gamma$ is opposite to lead $\gamma$.  We find a similar expression for the charge current, which coincides with the results of Ref.~\onlinecite{swirkowiczPRB06}, wherein the equilibrium contribution vanishes (as it should):
\eq{
\label{c_curr}
J_\gamma = -\frac{2e}{h}\textnormal{Re}\int  d\omega\textnormal{Tr}\Bigg[\left[f_{\bar\gamma}(\omega)-f_\gamma(\omega)\right]\left\{\hat{W}^{R}_\gamma\hat{G}^R(\omega)\hat W^R_{\bar\gamma}\left(\hat W^R_l + \hat W^R_r\right)^{-1}\left[\hat G^R(\omega)^{-1}-\hat G^A(\omega)^{-1}\right]\hat{G}^A(\omega)\right\}\Bigg]\,.
}
Because Ng's ansatz does not guarantee continuity of current when the Green's functions are matrix valued, we symmetrize the steady-state charge and spin currents as $J=(J_l-J_r)/2$ and $\textbf{J}=(\textbf{J}_l-\textbf{J}_r)/2$, respectively. Here, we focus our attention on the case with no external magnetic field, such that the isolated dot is SU(2) symmetric. 

In equilibrium, $f_{l}(\omega)=f_{r}(\omega)=f_{0}(\omega)$, the spin current can be written down exactly from Eq.~\eqref{neq_curr} by invoking the fluctuation-dissipation theorem, $\hat{G}^<(\omega)=f_{0}(\omega)[\hat{G}^A(\omega)-\hat{G}^R(\omega)]$:
\eq{
\textbf{J}_\gamma=-\mbox{Re}\int\frac{ d\omega}{2\pi} \textnormal{Tr}\Bigg\{f_{0}(\omega)\hat{W}_\gamma^{R}\hat{\boldsymbol{\sigma}} \left[\hat G^R(\omega) + \hat G^A(\omega)\right]-\hat W^{R}_\gamma \hat{\boldsymbol{\sigma}}\frac{i}{\pi}\mP\int d\omega' f_{0}(\omega')\frac{\hat{G}^A(\omega')-\hat{G}^R(\omega')}{\omega'-\omega}\Bigg\}\,.
\label{equil_cur}
}
The magnetic leads are invariant under the composition of time reversal and spin-space reflection with respect to the $xz$ plane. Therefore, because the dot is isotropic, the only equilibrium current that can persist is $J^y$ while all others vanish, $J=J^x=J^z=0$. Using the same transformation, we argue that the total spin density on the dot along the $y$ direction must be zero.

\end{widetext}

\section{Green's Functions}
\label{green}

We now specialize the above formalism to calculation of the charge and spin current in two extreme regimes of the Anderson model: zero on-site charging energy (noninteracting dot) and infinite on-site dot charging energy, which, in the temperature regime considered below, leads to the onset of Kondo features. The Anderson Hamiltonian for a single level is
\eq{H_{D}=\epsilon\sum_{\sigma}n_{\sigma}+Un_{\uparrow}n_{\downarrow}\,,}
where $n_\sigma=d^\dagger_\sigma d_\sigma$ and $U$ parametrizes the strength of Coulomb repulsion on the dot: either zero or infinite.

\begin{figure}[pt]
\includegraphics[width=\linewidth,clip=]{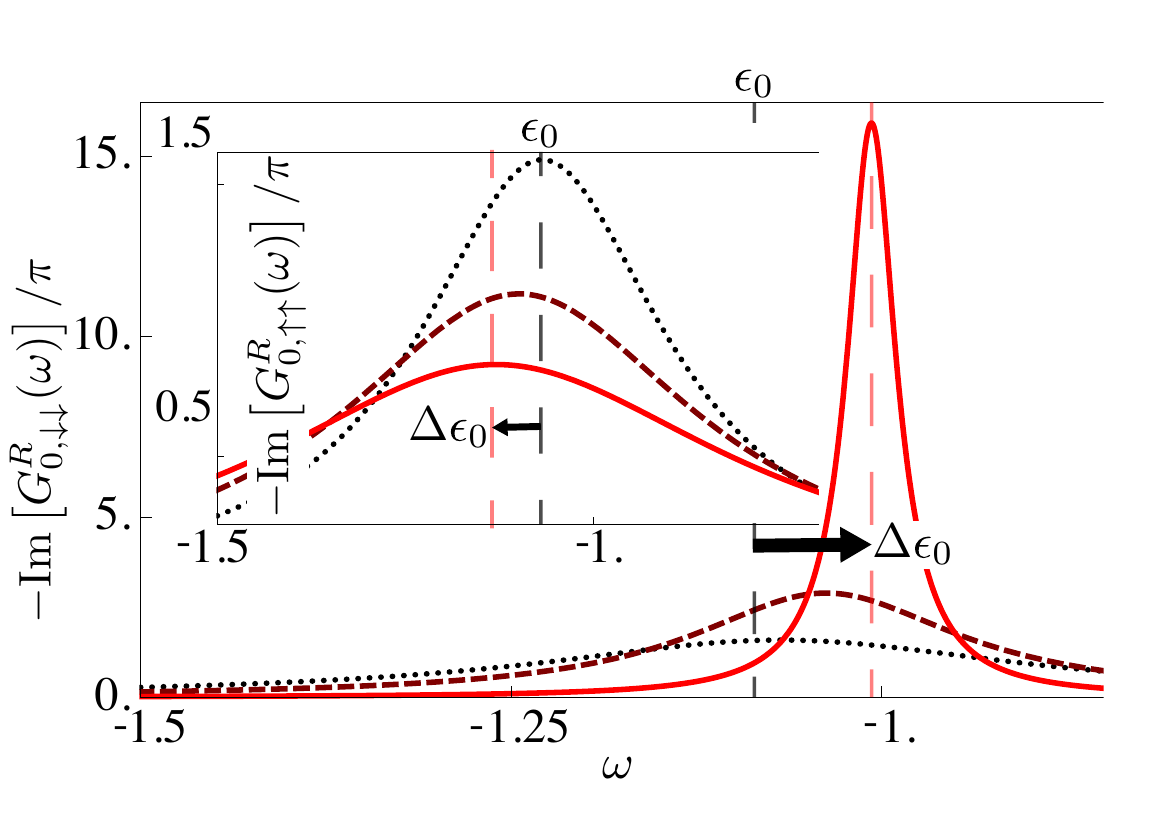}
\caption{The density of states for the spin-down electron of the noninteracting dot at $\theta=0$ (solid), $2\pi/3$ (dashed), and $\pi$ (dotted) with parameters $\epsilon=-1$, $\Gamma=0.2$, $\alpha=0$, $p_l=p_r=0.9$, $3\mathcal{W}_h=\mathcal{W}_e=75$, and $\theta_l=-\theta_r=\theta/2$. Inset: spin-up density of states of the noninteracting dot with the same parameters. There is an overall shift of the average energy resonance from $\epsilon$ to $\epsilon_0$ while the vertical dashed lines indicate a shift of $\Delta\epsilon_0$ above (below) $\epsilon_0$ for the spin-down (up) density of states.}
\label{DOSf}
\end{figure}

Using the equation of motion technique,\cite{haugBK08} one finds the noninteracting retarded Green's function to be $\hat{G}^R_0=(\omega-\epsilon-\hat{\Sigma}_0)^{-1}$, where\cite{rudzinskiPRB05}
\eq{\hat\Sigma_0=\left(\hat W_l^R+\hat W_r^R\right)\left(1+\frac{1}{i\pi}\ln\left|\frac{\omega-\mathcal{W}_e}{\omega+\mathcal{W}_h}\right|\right)\,.
\label{sig0}
}
Here, $\mathcal{W}_e$ ($\mathcal{W}_h$) is the electron bandwidth above (below) the chemical potential in the leads, and we have assumed that the biases in the leads are much smaller than the bandwidths, i.e., $\mu_{\sigma\gamma}\ll\mathcal{W}_e,\mathcal{W}_h$, in order to recover Eq.~(\ref{sig0}) out of equilibrium. Because the relevant energy scales are in general much smaller than the bandwidth, we take $\hat\Sigma_0$ to be independent of $\omega$. When the bandwidth is particle-hole asymmetric, i.e., $\mathcal{W}_e\neq \mathcal{W}_h$, the real part of the self-energy both renormalizes the bare dot resonance by a shift proportional to $\Gamma$ as well as lifts spin degeneracy by $\sim \sum p_\gamma\Gamma_\gamma\cos\theta_\gamma$. If $\mathcal{W}_h=\mathcal{W}_e$, the real part of the self-energy vanishes at $\omega=0$. Since the dot is noninteracting, $\hat\Sigma_0$ is independent of the position of the equilibrium chemical potential. In Fig.~\ref{DOSf}, we plot spin-dependent densities of states for $3\mathcal{W}_h=\mathcal{W}_e=75$ (in units of $|\epsilon|$), $\epsilon=-1$, $\Gamma=0.2$, $\alpha=0$, $p_l=p_r=0.9$, which show zero spin splitting when the leads are antiparallel (dotted curve), maximum effective spin splitting when the leads are parallel (solid curve), and an intermediate splitting between these values when $\theta_l=-\theta_r=\pi/3$ (dashed curve). Notice the average of the spin-up and down resonances is shifted from $\epsilon$ to $\epsilon_0\approx-1.1$. Furthermore, the spin degeneracy is broken so that the spin-down (up) resonance is shifted above (below) $\epsilon_0$ by $\Delta\epsilon_0\approx0.07$ as indicated by the vertical dashed lines in Fig.~\ref{DOSf} (Fig.~\ref{DOSf} inset). Interchanging $\mathcal{W}_e\leftrightarrow\mathcal{W}_h$, while keeping all other parameters fixed, both the overall shift in the dot resonance as well as the effective spin splitting reverse sign. When $\mathcal{W}_e=\mathcal{W}_h$, $\textrm{Re}\hat\Sigma_0=0$ and the resonance for both bands is at $\epsilon$, independent of the magnetic orientations in the leads.

When the Coulomb repulsion is large, the proximity-induced\cite{braunPRB04} spin splitting on the dot can be understood according to the Schrieffer-Wolff transformation.\cite{schriefferPR66} By treating the lead electrons as a mean field, to lowest order in $\Gamma$, the magnetic lead $\gamma$ splits the energy degeneracy on the dot by\cite{limPRB13}
\eq{
\Delta\epsilon_\gamma=\sum_{\sigma}\sigma\Gamma_{\sigma\gamma}\mathcal{P}\int\frac{d\varepsilon}{\pi}\left[\frac{1-f_{\sigma\gamma}(\varepsilon)}{\varepsilon-\epsilon}+\frac{f_{\sigma\gamma}(\varepsilon)}{\varepsilon-\epsilon-U}\right]
\label{de}
}
along the lead magnetic orientation, $\textbf{n}_\gamma=(\cos\theta_\gamma,\sin\theta_\gamma)$, where $\sigma=\pm$ for spin up/down. Unless otherwise specified, we choose $\theta_l$ and $\theta_r$ (for a given misalignment $\theta\equiv\theta_l-\theta_r$) in this and the following sections such that the total effective splitting $\Delta\epsilon=\sum_\gamma\Delta\epsilon_\gamma\cos\theta_\gamma$ is along the $z$ axis: $H_D=\sum_\sigma (\epsilon-\sigma\Delta\epsilon/2)n_\sigma+Un_\uparrow n_\downarrow$. When $U=0$, the spin splitting in Eq.~\eqref{de} reduces to the above noninteracting case. We henceforth assume $\mathcal{W}_e=\mathcal{W}_h$ so that $\hat\Sigma_0$ is purely imaginary and the up and down resonances are split only in the interacting case.

While the interacting Green's function cannot be found exactly, using the equation-of-motion method and truncating higher-order correlations within the leads,\cite{meirPRL93} the retarded Green's function, when $U\rightarrow\infty$, is given by $\hat{G}^R=(\omega-\epsilon-\hat{\Sigma}_0+\Delta\epsilon\hat\sigma^z/2-\hat\Sigma')^{-1}\hat{m}$, where\cite{swirkowiczPRB06}
\begin{align}
\Sigma'_{\sigma\sigma} &= \sum_{\sigma'\gamma}\int \frac{d \varepsilon}{2\pi}\frac{f_{\sigma'\gamma}(\varepsilon-\sigma\Delta\epsilon)\rho_{\sigma'\gamma}(\varepsilon)V_{\sigma\sigma'\gamma} V^*_{\bar{\sigma}\sigma'\gamma}}{\omega + i\eta - \varepsilon}\,,\nonumber\\
\Sigma'_{\sigma\bar{\sigma}}
& = - \sum_{\sigma'\gamma}\int \frac{d \varepsilon}{2\pi}\frac{f_{\sigma'\gamma}(\varepsilon)\rho_{\sigma'\gamma}(\varepsilon)V_{\sigma\sigma'\gamma} V^*_{\bar{\sigma}\sigma'\gamma}}{\omega + i\eta - \varepsilon}\,,
\label{S1}
\end{align}
$m_{\sigma\sigma}=1-\av{n_{\bar{\sigma}}}$, $m_{\sigma\bar{\sigma}}=\av{d_{\bar{\sigma}}^\dagger d_\sigma}$, $\rho_{\sigma\gamma}$ is the spin-dependent density of states in lead $\gamma$, $\bar\sigma=-\sigma$, and $\eta\rightarrow0^+$. The density-matrix elements must be calculated self-consistently according to
\eq{
\av{n_\sigma}=\mbox{Im}\int\frac{ d\omega}{2\pi}G^<_{\sigma\sigma}(\omega)\,,\,\,\,\av{d_{\bar{\sigma}}^\dagger d_\sigma}=-i\int\frac{d\omega}{2\pi}G^<_{\sigma\bar{\sigma}}(\omega)\,.
}  
The details of the calculation of the Kondo Green's function can be found in appendices \ref{lgf} and \ref{agf}. In addition to the ordinary broadening of single-particle levels due to tunneling [see Eq.~\eqref{sig0}], which we treat exactly, we take, to lowest order in tunneling, processes corresponding to spin flips in the usual Kondo problem,\cite{hewsonBK97} which lead to logarithmic divergences.

\begin{figure}[pt]
\includegraphics[width=\linewidth,clip=]{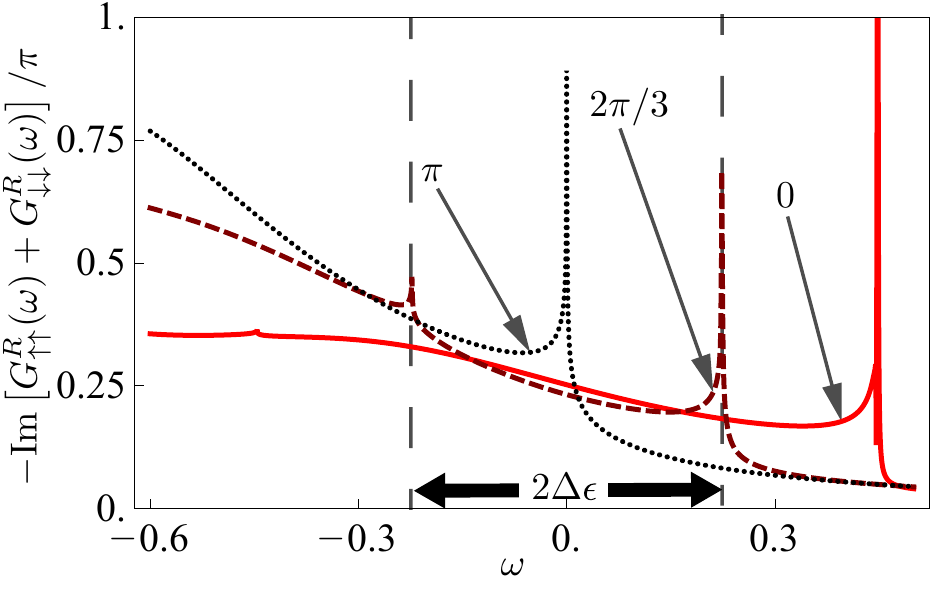}
\caption{Equilibrium Kondo density of states of the dot at $\theta=0$ (solid), $2\pi/3$ (dashed), and $\pi$ (dotted), setting $\mu=0$, $\epsilon=-1$, $\Gamma=0.2$, $\alpha=0$, and $p_l=p_r=0.9$. By the vertical dashed lines, we indicate a splitting of the Kondo peak due to the effective exchange field of $2\Delta\epsilon\approx0.5$ for the specific case of $\theta=2\pi$/3.}
\label{DOSK}
\end{figure}

This simplified treatment of the Anderson model captures qualitative aspects of the Kondo physics and should give us guidance regarding the position of singularities. Because the sharpness of the Fermi sea gives rise to the Kondo peaks, we expect a resonance for each Fermi surface in the leads. In the general case of noncollinear magnetic leads with spin-dependent biases, there could be, according to Eqs.~(\ref{S1}), eight peaks at $\mu_{\uparrow l}\pm\Delta\epsilon$, $\mu_{\downarrow l}\pm\Delta\epsilon$, $\mu_{\uparrow r}\pm\Delta\epsilon$, and $\mu_{\downarrow r}\pm\Delta\epsilon$ from the diagonal of $\hat\Sigma'$ and four at $\mu_{\uparrow l}$, $\mu_{\downarrow l}$, $\mu_{\uparrow r}$, and $\mu_{\downarrow r}$ from the off-diagonal of $\hat\Sigma'$. When the bias is spin independent, $\mu_{\uparrow\gamma}=\mu_{\downarrow\gamma}$, the number of Kondo peaks is halved to six at $\mu_l$, $\mu_r$, $\mu_l\pm\Delta\epsilon$, and $\mu_r\pm\Delta\epsilon$. In equilibrium, one can show [according to the form \eqref{hV} of the tunneling matrix elements] that $\Sigma'_{\sigma\bar\sigma}=0$, and, therefore, there are only two Kondo peaks near the Fermi surface shifted by the exchange-proximity field, $\mu\pm\Delta\epsilon$, as seen in the equilibrium density of states plotted in Fig.~\ref{DOSK}, taking $\mathcal{W}_e=\mathcal{W}_h=50$ while keeping all other parameters as before. As a specific example, when $\theta=2\pi/3$, the Kondo peaks are split by $2\Delta\epsilon\approx0.5$. Notice that the single peak is restored when the leads are oriented antiferromagnetically, i.e., in the absence of an effective exchange splitting.

The decoupling scheme used to calculate the Kondo Green's function is applicable when the temperature in the leads is above the Kondo temperature, $T_l,T_r\gtrsim T_K\sim \sqrt{\Gamma\mathcal{W}}\exp(-\pi\left|\mu-\epsilon\right|/\Gamma)$.\cite{meirPRL93} Below $T_K$, some of the correlators that are assumed to vanish in our approximation (see appendix~\ref{agf}) start diverging. Within the framework of our analysis, a finite temperature introduces a broadening of the Kondo singularity. For the purpose of our discussion, we, nonetheless, took $T=0$, which should not qualitatively affect our main conclusions.

\section{Equilibrium}
\label{equilibrium}

In the absence of a nonequilibrium bias, i.e., $\mu_l=\mu_r=\mu$ and $T_l=T_r$, the symmetry of our setup allows only for spin current polarized along the $y$ axis and dot polarization along the $z$ axis. In this and the following sections, we consider a dot symmetrically coupled ($\alpha=0$ and $p_l=p_r\equiv p$) to the leads, where $\epsilon=-1$, the bandwidth in the leads is $\mathcal{W}=2\mathcal{W}_h=2\mathcal{W}_e=100$, and $k_B T=10^{-3}$, setting $\mu=0$. (The energy is measured in units of the detuning energy $\Delta=\left|\mu-\epsilon\right|$.)

\begin{figure}[pt]
\includegraphics[width=\linewidth,clip=]{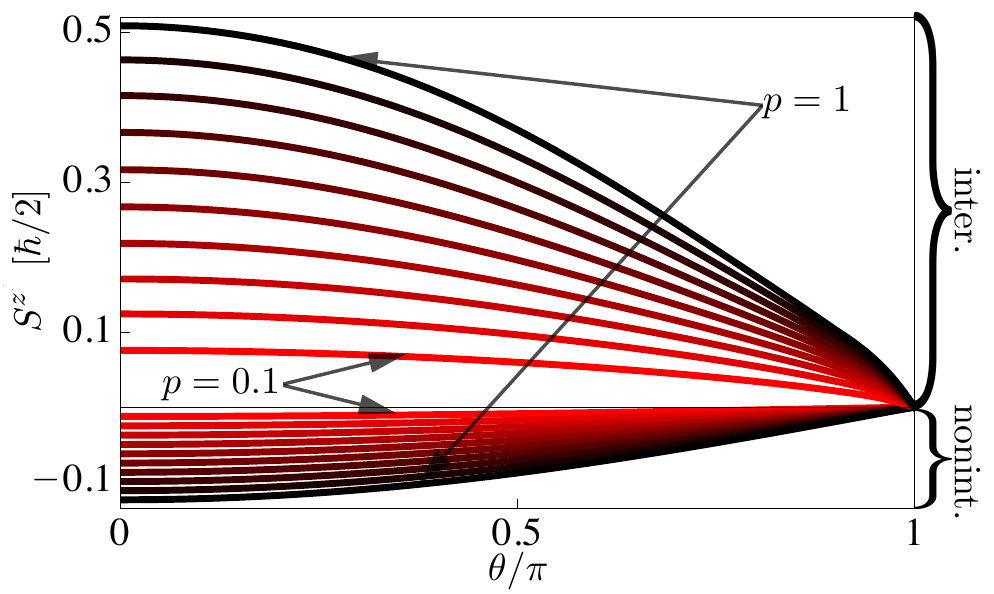}
\caption{Spin polarization of the dot as a function of the relative canting of the magnetic leads for several values of lead polarization, $p=0.1,0.2,\dots,1$, and $\Gamma=0.2$. The positive (negative) values of spin are for the interacting (noninteracting) dot.}
\label{eq_Sz}
\end{figure}

We plot the polarization of the noninteracting and the interacting dots (Fig.~\ref{eq_Sz}) as a function of $\theta$ for $\Gamma=0.2$ and several values of lead polarizations varying from $p=0.1$ to $p=1$, increasing in increments of $0.1$. For both the noninteracting and interacting dots, the magnitude of the dot polarization is maximized when the magnetization directions of leads is parallel and increases with lead polarization. The magnetic moment of the dot is antiparallel and parallel to the $z$ axis in the noninteracting and interacting cases, respectively, and the magnitude of polarization is about a factor of five larger in the interacting case, due to the large on-site repulsion, when $p=1$. 
Although one may na\"ively expect the magnetic moment of the noninteracting dot to point along the exchange field (corresponding to the majority spins in the leads), in the parallel alignment case, the stronger tunneling in the spin-up sector hybridizes the electrons on the leads and dot, delocalizing the up spin, which leaves a net antiparallel polarization on the dot when $\epsilon<0$. If the dot resonance was above the Fermi level in the leads, $\epsilon>0$, the net polarization on the dot would be parallel to the exchange field, according to the particle-hole symmetry.

\begin{figure}[pt]
\includegraphics[width=\linewidth,clip=]{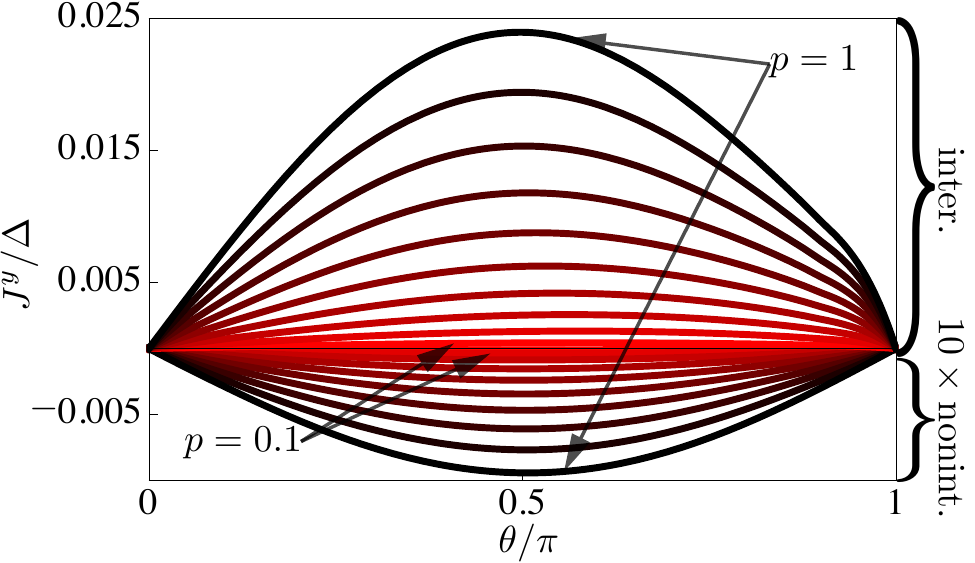}
\caption{$y$ component of the spin current flowing from left to right, for a noninteracting dot as a function of the relative canting of the magnetic leads for several values of lead polarization, $p=0.1,0.2,\dots,1$, and $\Gamma=0.2$.}
\label{eq_Jy}
\end{figure}

The spin current, similarly, is negative (for both $\epsilon=\pm1$) and small for the free dot but positive and relatively large for the interacting dot (Fig.~\ref{eq_Jy}). This means that the magnetic leads separated by a quantum dot with resonance below the Fermi surface tend to have a weak antiferromagnetic exchange coupling when there is no interaction on the dot, whereas, when there is a large on-site repulsion, the magnets are subject to a stronger ferromagnetic coupling (which bears some analogy with the double-exchange mechanism).

\begin{figure}[pt]
\includegraphics[width=\linewidth,clip=]{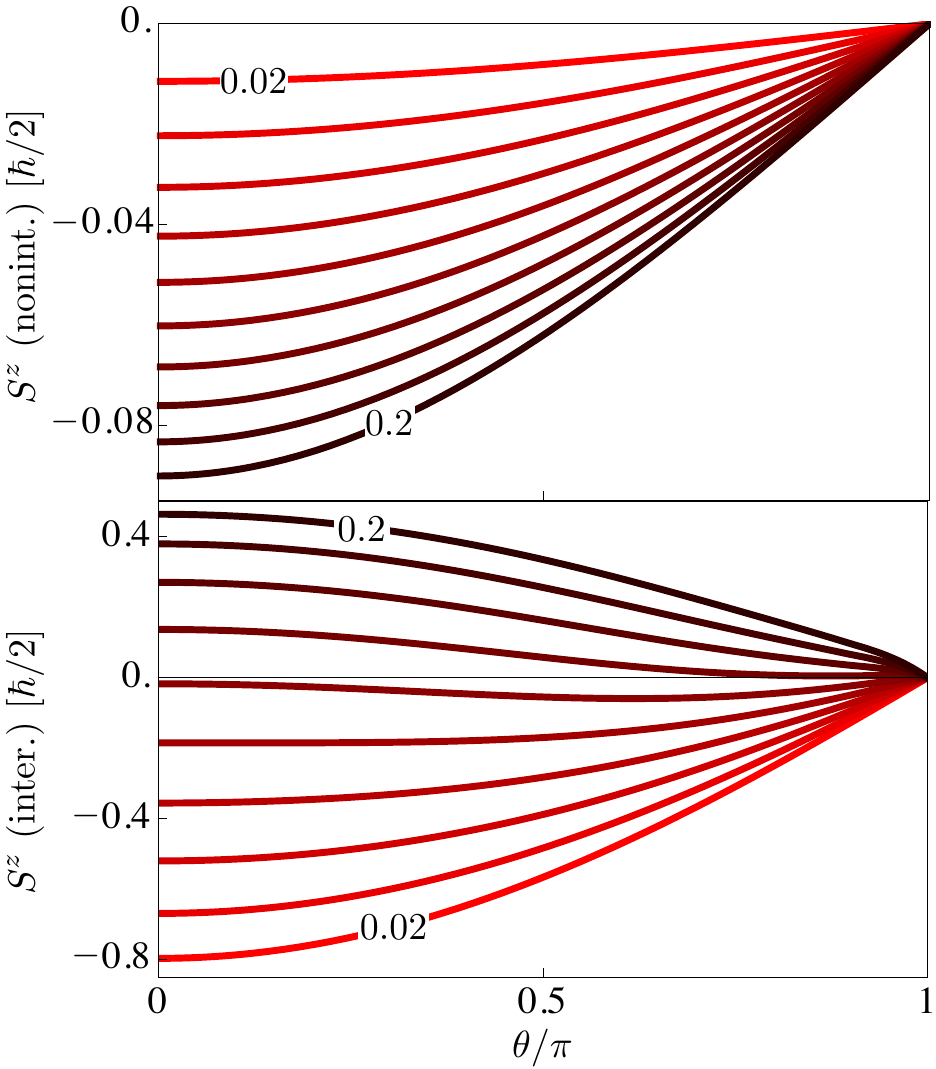}
\caption{Spin polarization of the interacting (top) and noninteracting (bottom) dot as a function of the relative canting of the magnetic leads for $\Gamma=0.02,\dots,0.2$, in increments of $0.02$, and $p=0.9$.}
\label{eq_Sz_gam}
\end{figure}

In Fig.~\ref{eq_Sz_gam}, we plot the average spin on the dot for several values of $\Gamma$, taking the same parameters as before while fixing polarization $p=0.9$. When the dot is noninteracting, the polarization monotonically decreases as the dot decouples from the leads. For large on-site repulsion, the spin of the nearly decoupled dot is antiparallel to the exchange field and decreases with increasing $\Gamma$. Increasing the coupling beyond some critical value, $\Gamma\approx0.1$, the polarization passes through zero and becomes positive. That is, when the coupling is weak, there are no Kondo correlations and the system is a singly-occupied quantum dot with spin that approaches $-\hbar/2$. As the coupling is increased, many-body correlations between the dot and the leads build up reversing the spin polarization of the dot.

\begin{figure}[pt]
\includegraphics[width=\linewidth,clip=]{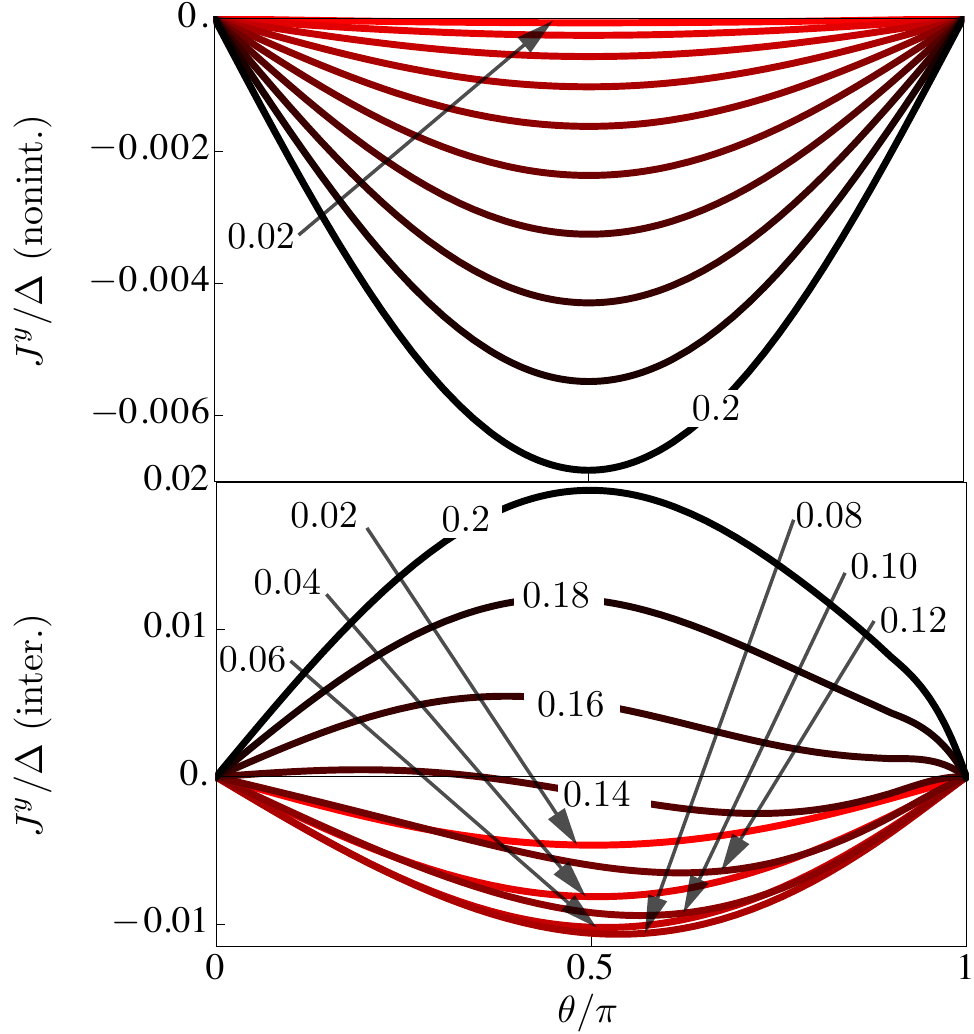}
\caption{Equilibrium spin current of the noninteracting (top) and interacting (bottom) dot, as a function of the relative canting of the magnetic leads for $\Gamma=0.02,\dots,0.2$, in increments of $0.02$, and $p=0.9$.}
\label{eq_Jy_gam}
\end{figure}

\begin{figure}[pt]
\includegraphics[width=\linewidth,clip=]{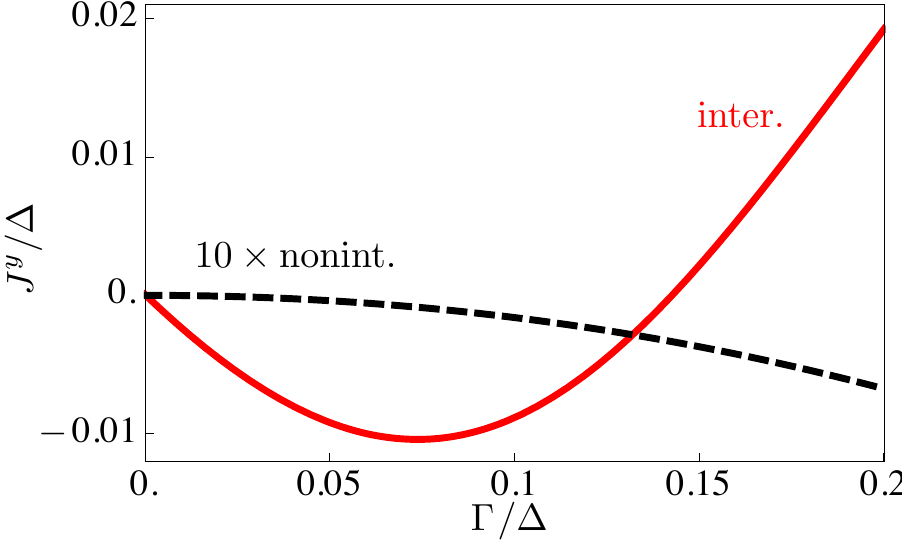}
\caption{Equilibrium spin current of the interacting (solid) and noninteracting (dashed) dot as a function of $\Gamma$ for canted magnetic leads fixed at $\theta=\pi/2$ for $p=0.9$.}
\label{eq_Jy_var_gam}
\end{figure}

Similarly to the spin polarization, the spin current for the noninteracting dot monotonically increases with increasing coupling. See Fig.~\ref{eq_Jy_gam} (upper panel). The interacting dot [Fig.~\ref{eq_Jy_gam} (lower panel)], on the other hand, exhibits a nonmonotonic behavior, with the spin current changing sign as $\Gamma$ is increased. This is shown in Fig.~\ref{eq_Jy_var_gam}, where we plot the spin current as a function of $\Gamma$ at $\theta=\pi/2$. 

\section{Voltage Biasing}
\label{voltage}

Next, we symmetrically apply a (spin-independent) voltage difference to the leads, $\mu_l=-\mu_r=eV/2$, and calculate the differential conductance, $G=dJ/dV$, and the differential spin conductance, $\textbf{G}=d\textbf{J}/dV$. (Throughout this section, we set $\epsilon=-1$, $\Gamma=0.2$, $\alpha=0$, and $p_l=p_r=0.9$.) In this setup, under a $\pi$ rotation (both in real and spin space) around the $z$ axis, the voltage difference as well as the charge current and $z$ component of spin current reverse sign, whereas the $x$ and $y$ components of spin current are invariant. Accordingly, as a function of voltage difference, $G$ and $G^{z}$ ($G^{x}$ and $G^y$) must be symmetric (antisymmetric).  

\begin{figure}[pt]
\includegraphics[width=\linewidth,clip=]{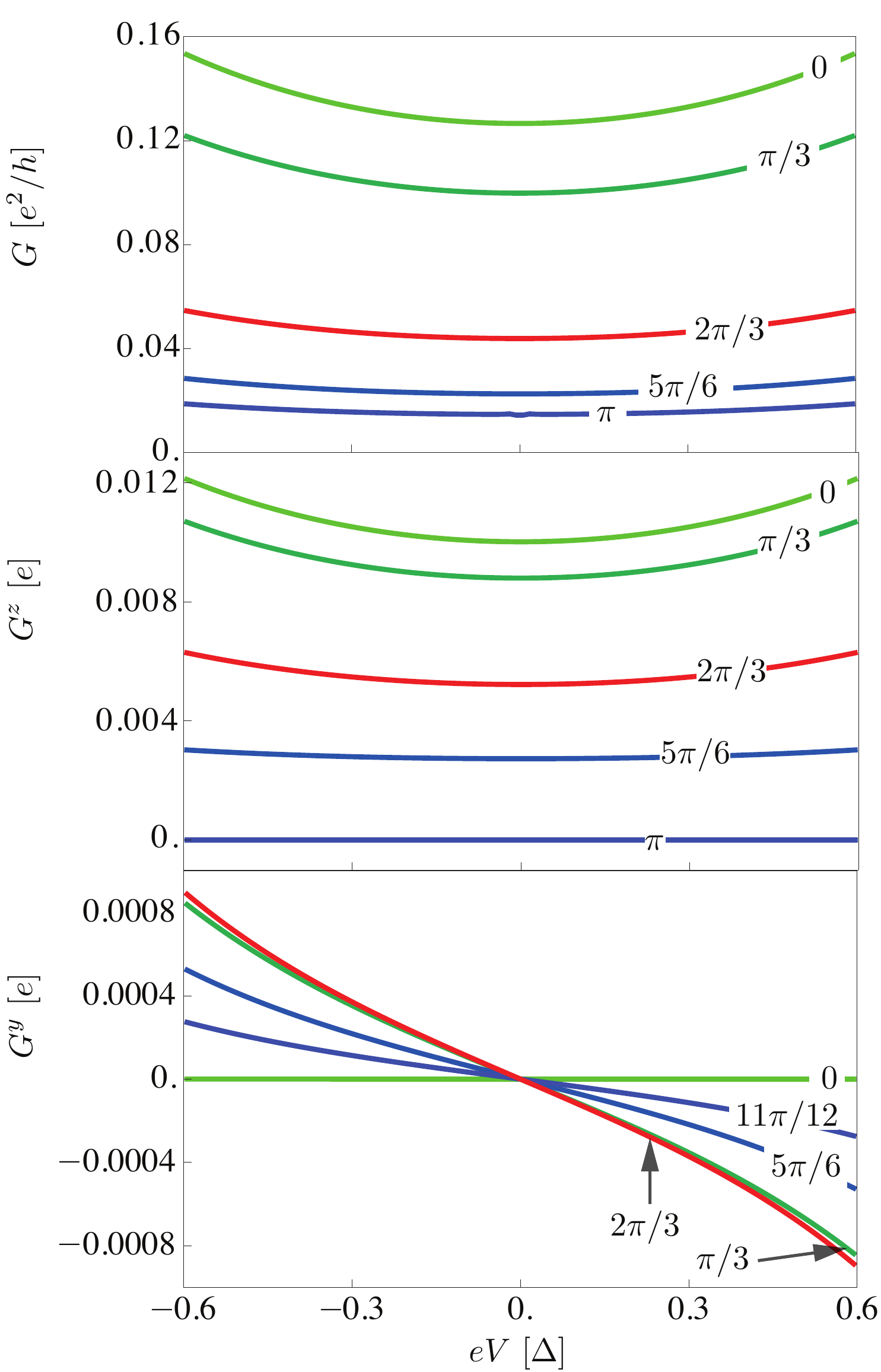}
\caption{Differential conductance (top panel), $z$ component (middle panel), and $y$ component (bottom panel) of the spin differential conductance for the noninteracting dot as a function of bias voltage $eV$ for several values of $\theta$.}
\label{Gall_f}
\end{figure}

We plot the differential conductance of the free dot for several values of $\theta$ in Fig.~\ref{Gall_f} (top panel). When there is no on-site interaction, the conductance is flat near zero bias and increases as the voltage approaches the resonance of the dot, $\epsilon=-1$. The decrease in conductance as the relative angle of the leads is increased from $0$ to $\pi$ reflects the mismatch of the density of states between up and down conduction bands in the leads, which is responsible for the tunnel magnetoresistance.\cite{jullierePLA75} In the antiparallel configuration near zero bias, there is a small dip with width proportional to the temperature in the leads: When $eV\lesssim k_B T$, the current is smeared over an energy range proportional to the temperature, around the Fermi level.
 
The $z$ component of the spin differential conductance [Fig.~\ref{Gall_f} (center panel)] is similarly symmetric in voltage and decreases as the magnetic misalignment of the leads is increased. $G^y$ [Fig.~\ref{Gall_f} (bottom panel)] is, consistently with symmetry, monotonically decreasing as a function of voltage difference. It is exactly zero when the leads are parallel and largest in magnitude when the magnetization vectors are approximately perpendicular.

\begin{figure}[pt]
\includegraphics[width=\linewidth,clip=]{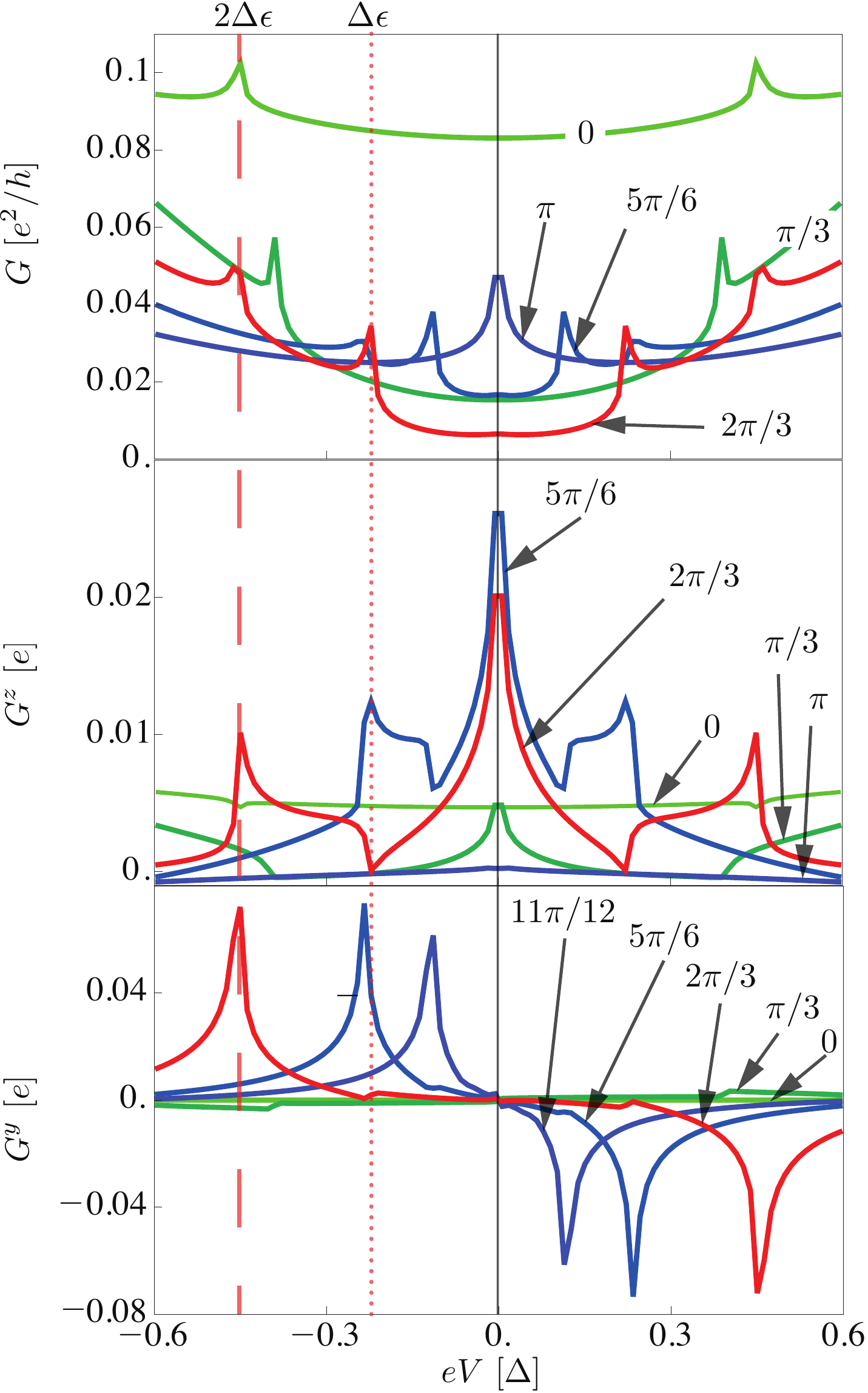}
\caption{Differential conductance (top panel), $z$ component (middle panel), and $y$ component (bottom panel) of the spin differential conductance for the interacting dot as a function of bias voltage $eV$ for several values of $\theta$. For $\theta=2\pi/3$, the dotted (dashed) vertical line indicates features in the conductance at $eV=\Delta\epsilon$ ($eV=2\Delta\epsilon$).}
\label{Gall_K}
\end{figure}

\begin{figure}[pt]
\includegraphics[width=\linewidth,clip=]{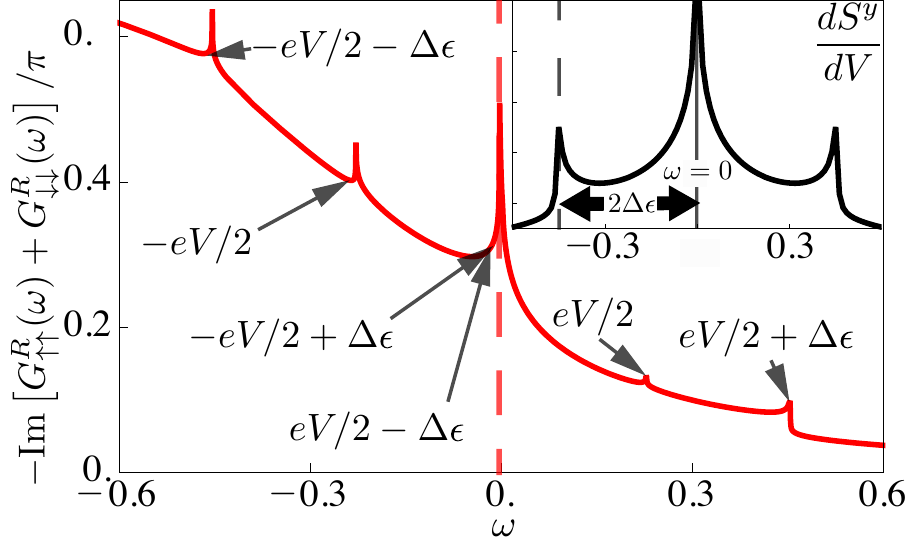}
\caption{Density of states of the interacting dot for  $\theta=2\pi/3$ when $eV=2\Delta\epsilon\approx0.5$, wherein we point out the positions of each of the six expected Kondo resonances; note the overlap of the resonances at $eV/2-\Delta\epsilon=-eV/2+\Delta\epsilon$ indicated by the dashed vertical line at $\omega\approx0$. Inset: Derivative of the spin polarization of the dot in the $y$ direction. We indicate by the vertical dashed line a peak at $eV=2\Delta\epsilon$.}
\label{mu_DOS}
\end{figure}

For the interacting dot, peaks in conductance occur when the difference in chemical potentials between the two leads is equal to the exchange splitting, $\Delta\epsilon$.\cite{swirkowiczPRB06} See Fig.~\ref{Gall_K} (upper panel). Because the Kondo peaks are shifted away from the chemical potential when the leads are not antiparallel, the Kondo resonance is inaccessible at zero bias. However, when $eV=\Delta\epsilon$, the Kondo resonances at $\mu_l-\Delta\epsilon$ and $\mu_r+\Delta\epsilon$ become energetically accessible and there is a peak in conductance corresponding to the additional transport through these channels.\cite{meirPRL93} There are an additional two peaks when the difference in chemical potentials between the left and right leads is equal to twice the exchange splitting. Recall that for the spin-independent bias there are, generally, six Kondo peaks; when $eV=2\Delta\epsilon$ the resonances at  $eV/2-\Delta\epsilon$ and $-eV/2+\Delta\epsilon$ are both positioned at zero, thus partially restoring the magnetically unperturbed zero-bias peak. Because the spin and charge currents, Eqs.~\eqref{s_curr} and \eqref{c_curr}, are nonlinear in the Green's functions, we indeed expect some feature (in this case a peak) at $eV=2\Delta\epsilon$. For the particular example of $\theta=2\pi/3$, we point out in Fig.~\ref{Gall_K} the peaks in conductance at $eV=\Delta\epsilon\approx0.25$ (dotted vertical line) and $eV=2\Delta\epsilon\approx0.5$ (dashed vertical line).  Recalling Fig.~\ref{DOSK}, the peaks in conductance at $eV\approx\pm0.25$ are at the same energy as the Kondo resonances in the equilibrium density of states. On the other hand, when $eV=2\Delta\epsilon\approx0.5$, the nonequilibrium density of states (Fig.~\ref{mu_DOS}) yields a partially restored zero-bias peak. This manifests in the density matrix: $d S^y/dV$ as a function of $eV$ shows a peak at precisely $2\Delta\epsilon$ [Fig.~\ref{mu_DOS} (inset)]. Because the retarded Green's function depends on the density matrix, there is a peak in conductance at $2\Delta\epsilon$. 

The $z$ component of the spin conductance [Fig.~\ref{Gall_K} (center panel)], likewise, is enhanced at $eV=2\Delta\epsilon$, however, it is diminished at $\Delta\epsilon$. At $eV=\Delta\epsilon$, the newly accessible resonances at $\mu_l-\Delta\epsilon$, corresponding to the majority band, and $\mu_r+\Delta\epsilon$, corresponding to the minority band, contribute oppositely in spin current along the $z$ axis, resulting in a suppression of $G^{z}$. On the other hand, at $eV=2\Delta\epsilon$, two Kondo peaks align (cf. Fig.~\ref{mu_DOS}) in the density of states, and the nonlinearities in the current cause an enhancement of the spin conductance, analogously to the case of charge conductance. There is a zero-bias peak in the spin conductance, when the leads are canted, which is rooted in the off-diagonal elements of the interacting self-energy $\hat{\Sigma}'$ (see Sec.~\ref{green}). The same holds for the zero-bias peak in the derivative of the $y$ component of dot polarization [cf. Fig.~\ref{mu_DOS} (inset)].

$G^y$ [Fig.~\ref{Gall_K} (bottom panel)] similarly has features at $eV=0$,~$\pm\Delta\epsilon$, and $\pm2\Delta\epsilon$. There are large enhancements when the bias is equal to twice the exchange splitting, small decreases at $\pm\Delta\epsilon$, and a small feature at zero bias. At the peaks, the $y$ component of the spin conductance is enhanced by two orders of magnitude in comparison to the noninteracting case. 

The differential conductance of spin current polarized along the $x$ axis (omitted from Figs.~\ref{Gall_f} and \ref{Gall_K}) is, analogously to $G^y$, antisymmetric as a function of voltage in both the noninteracting and interacting regimes, and has features at $eV=0,\pm\Delta\epsilon,\pm2\Delta\epsilon$.

\section{Microwave Precession}
\label{spin}

As another application of our formalism, we now consider spin and charge pumping induced by circular precession of the left magnetic lead at frequency $\Omega$, while taking the right lead to be nonmagnetic (i.e., setting $p_r=0$). Transforming to the frame of reference of the rotating magnetic lead, both leads are static but we must include a fictitious Zeeman splitting of $\hbar\Omega$ along the axis of precession and spin-dependent bias in the leads of\cite{tserkovPRB08} $\mu_{\uparrow(\downarrow) l}=\mp(\hbar\Omega/2)\cos\theta$ and $\mu_{\uparrow(\downarrow) r}=\mp\hbar\Omega/2$, where, departing from the convention of the previous sections, we fix $\theta_l=\theta$ (i.e., the angle of precession). The magnitude of the total spin splitting for the interacting dot is the magnitude of the vector sum of the fictitious and exchange fields, which are noncollinear: $\Delta\epsilon'=\sqrt{(\Delta\epsilon\sin\theta)^2+(\Delta\epsilon\cos\theta+\hbar\Omega)^2}$.  Taking all other parameters as in the previous section and zero applied voltage (which is now not necessary in order to drive nonequilibrium transport), we calculate the differential current response to microwaves, $G_s=dJ/d(\hbar\Omega)$, and the analogous quantity for the $z$-component of the spin current, $G^z_s=dJ^z/d(\hbar\Omega)$. We also find a finite spin current polarized along the $x$ and $y$ axes in the rotating frame of reference which, in the laboratory frame, contributes an ac spin current and averages to zero over a period of precession. Under a $\pi$ rotation around the $x$ axis in spin space, $\theta\rightarrow\pi-\theta$ and $\Omega\rightarrow-\Omega$ while $J\rightarrow J$ and $J^z\rightarrow -J^z$. From these geometric considerations,  $G_s(\Omega,\theta)=-G_s(-\Omega,\pi-\theta)$ and $G^z_s(\Omega,\theta)=G^z_s(-\Omega,\pi-\theta)$ for an arbitrary on-site interaction.

\begin{figure}[pt]
\includegraphics[width=\linewidth,clip=]{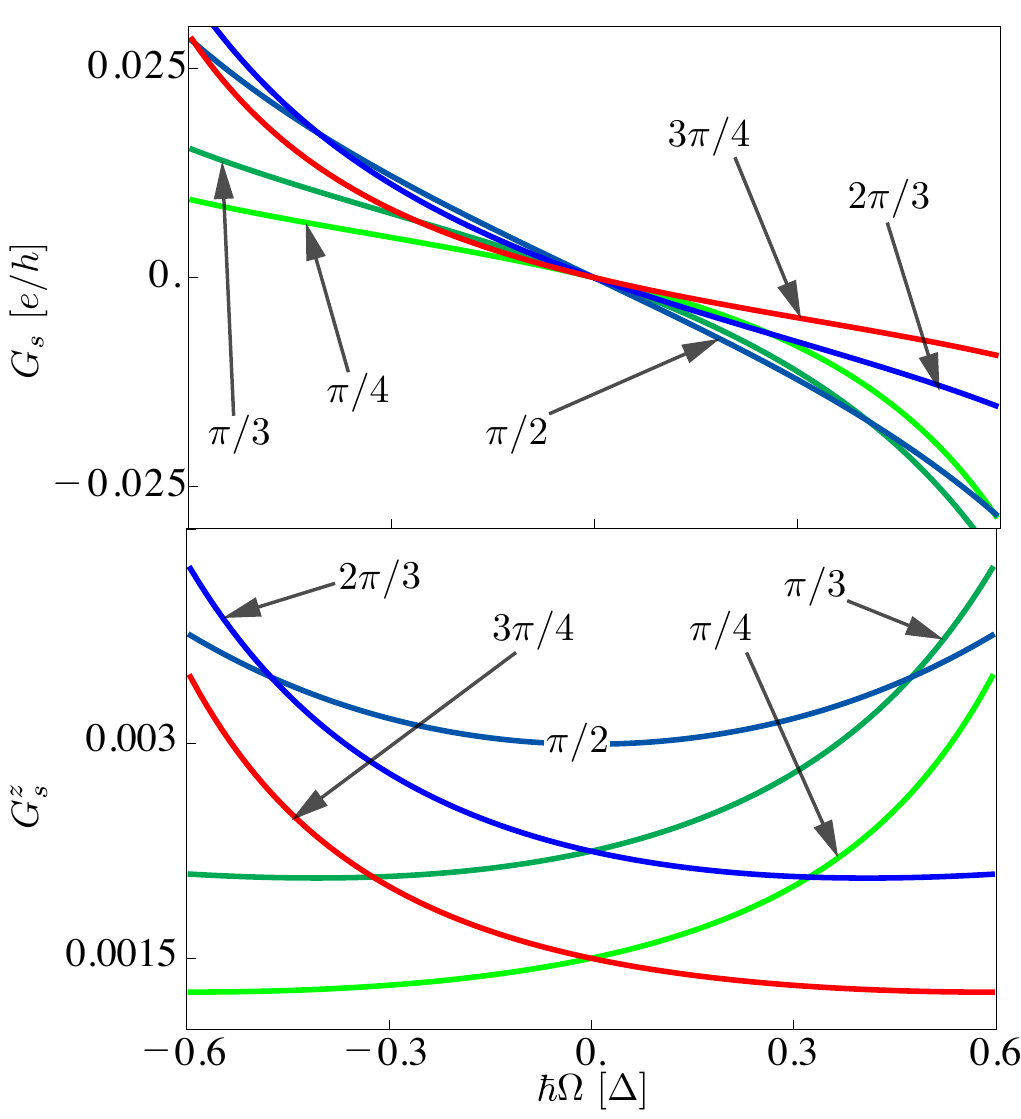}
\caption{Differential response of current (top panel) and spin current polarized along the $z$ axis (bottom panel) to microwaves of frequency $\Omega$ through the noninteracting dot  tunnel-coupled to a magnetic lead on the left and nonmagnetic lead on the right, for several angles of precession.}
\label{Gsall_f}
\end{figure}

\begin{figure}[pt]
\includegraphics[width=\linewidth,clip=]{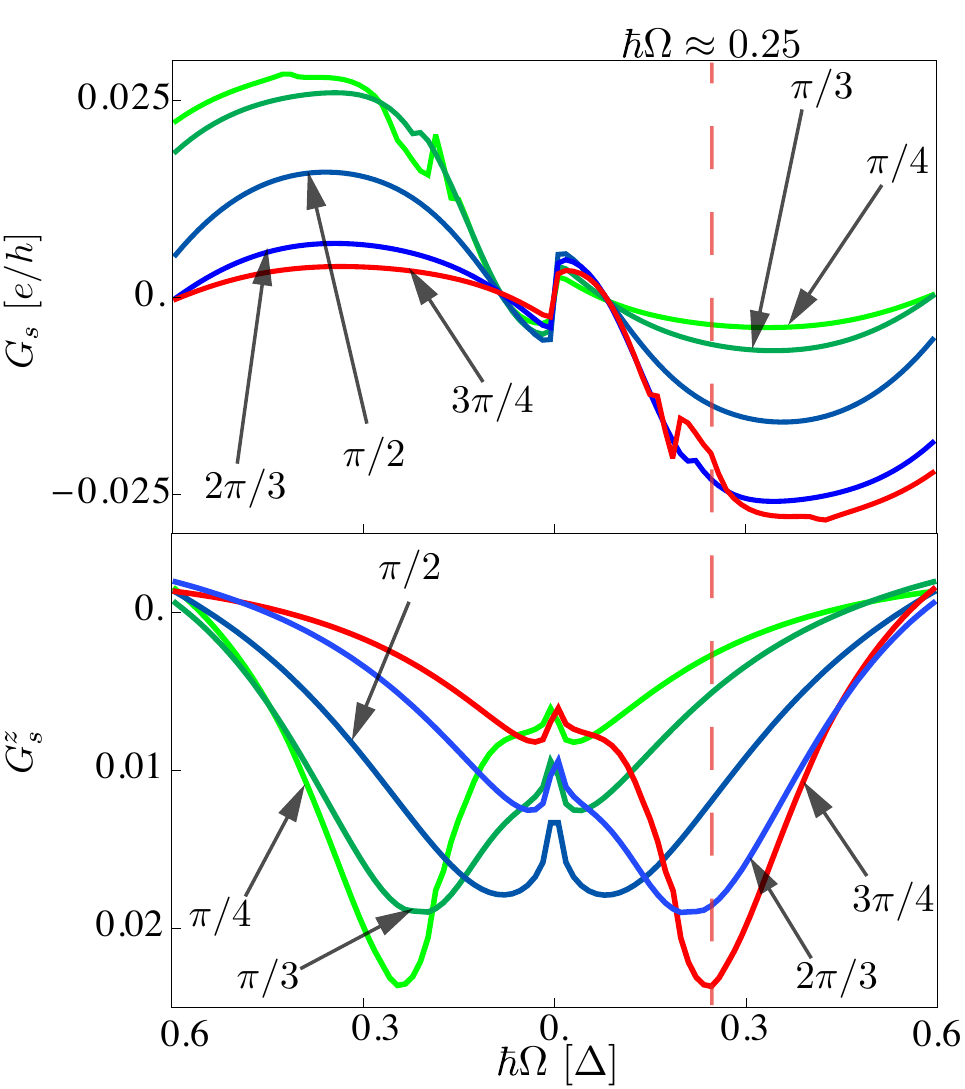}
\caption{Differential response of current (top panel) and spin current polarized along the $z$ axis (bottom panel) to microwaves of frequency $\Omega$ through the interacting dot  tunnel-coupled to a magnetic lead on the left and nonmagnetic lead on the right, for several angles of precession. We indicate by a dashed vertical line a feature in the current response at $\hbar\Omega\approx0.25$ as a result of the partial restoration of the Kondo resonance.}
\label{Gsall_K}
\end{figure}

For the noninteracting dot (Fig.~\ref{Gsall_f}), $G_s$ is exactly antisymmetric when $\theta=\pi/2$ and nearly antisymmetric for other values of $\theta$. We expect an enhancement of the transport when any of the spin-dependent chemical potentials in the leads is at the dot resonance for either up or down spin, $\epsilon\pm\hbar\Omega/2$. For an arbitrary $\hbar\Omega$, one resonance is always below the smallest chemical potential and therefore inaccessible. Increasing (decreasing) the microwave frequency, there is an upturn (downturn) in $G_s$ corresponding to the opening of the accessible transport channel.  When $\Omega\rightarrow-\Omega$, the spin up and down resonances, which in general have different spectral widths, exchange roles. Because $\left|\Sigma_{0,\downarrow\downarrow}\right|\neq\left|\Sigma_{0,\uparrow\uparrow}\right|$, the onset of the resonance occurs at different positions in the positive and negative-frequency regimes, hence some asymmetry in $G_s$. (There is no asymmetry when $\theta=\pi/2$ because $\Sigma_{0,\uparrow\uparrow}=\Sigma_{0,\downarrow\downarrow}$.) Using similar arguments, we explain the asymmetry in $G_s^z(\Omega)$.  

\begin{figure}[pt]
\includegraphics[width=\linewidth,clip=]{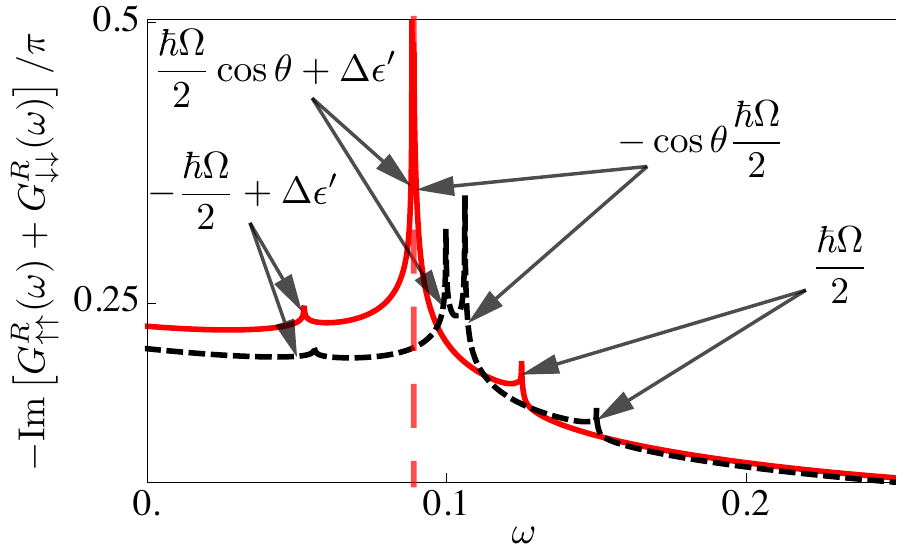}
\caption{Density of states of an interacting dot with magnetic lead precessing at angle $\theta=3\pi/4$ and frequency $\hbar\Omega\approx0.25$ (solid curve) and $\hbar\Omega=0.3$ (dashed curve). We point out the positions of the Kondo resonances in this energy range, specifically taking note of the overlap of the peaks at $-(\hbar\Omega/2)\cos\theta=(\hbar\Omega/2)\cos\theta+\Delta\epsilon'\approx0.08$ at precession frequency $\hbar\Omega\approx0.25$, indicated by the vertical dashed line.}
\label{spin_DOS}
\end{figure}

The microwave responses for the interacting dot are smooth for positive frequencies when $0<\theta<\pi/2$ and for negative frequencies when $\pi/2<\theta<\pi$. Conversely, when the frequency is positive and $\pi/2<\theta<\pi$ or the frequency is negative and $0<\theta<\pi/2$, the current response has several sharp features. Informed by the previous section, we expect these features to correspond to either the opening of transport channels or partial restoration of the Kondo resonances. The Kondo resonance is now split by both the fictitious magnetic field and spin-dependent splittings of the Fermi surfaces, both of which depend on the frequency of precession.  One may show\cite{Note1} that a necessary condition for partial reconstruction of the Kondo resonance, varying only $\Omega$, is $\Omega\cos\theta<0$, consistent with Fig.~\ref{Gsall_K}. For example, when $\theta=3\pi/4$, there is a visible feature in $G_s$ at $\hbar\Omega\approx0.25$ (dashed vertical line). In Fig.~\ref{spin_DOS}, we plot the density of states when $\theta=3\pi/4$ at frequency $\hbar\Omega\approx0.25$ (solid curve) and $\hbar\Omega=0.3$ (dashed curve). In the plotted range of energy, there are three distinct peaks for $\hbar\Omega\approx0.25$ and four for $\hbar\Omega=0.3$. The reduction in the number of peaks is due to the overlap of the resonances at $-(\hbar\Omega/2)\cos\theta=(\hbar\Omega/2)\cos\theta+\Delta\epsilon'\approx0.08$, i.e., a partial reconstruction of the original Kondo peak. Other features in the current response to microwaves may be explained analogously. 

\section{Summary and Outlook}
\label{summary}

In summary, correlation effects generically give rise to low energy peaks which manifest in both equilibrium and nonequilibrium spin currents. In equilibrium, this results in a characteristically larger spin current, as compared with the noninteracting case, as well as a nonmontonic dependence on the dot-lead tunneling rate. The exchange mediated by the spin current can therefore be either ferromagnetic or antiferromagnetic as a function of the tunneling strength. 

When driven by voltage, many-body correlations enhance both the charge and spin conductance as a direct result of the additional transport through conductance channels associated with the Kondo peaks. The enhancement is a manifestation of either the opening of additional transport or the partial reconstruction of Kondo peaks split by the effective exchange field from the leads. At specific low energy values of voltage, away from the dot resonance, the spin conductance can be increased by as a much as two orders of magnitude as compared with the noninteracting spin conductance.

Similarly, driving charge and spin transport by microwave precession enhances the charge and spin conductance. However, in contrast to voltage driving, the total effective exchange splitting on the dot and the bias dependence are both simultaneously dependent on the frequency of magnetic precession. We have shown that sharp features in the conductance, associated with the position of the Kondo resonances, may only occur when the projection of the exchange field along the axis of magnetic precesision is negative.

Although our calculations have been done for a single dot, one may consider a tunnel junction wherein impurities may be treated as an ensemble of quantum dots. Upon averaging over this ensemble, the characteristics of noninteracting dots in the spin current will be smeared out while preserving the signature of the interacting spin current induced by low energy peaks. As such, introducing impurities in tunnel junctions can enhance charge and spin current at sufficiently low temperatures.

Because correlation effects are responsible for these striking features at low energies, the current profiles, in turn, may help to elucidate many-body physics on quantum dots and in tunnel junctions. Therefore, charge and spin conductance measurements driven by voltage, temperature, or magnetic precession could serve as a natural probe of correlated systems.

\acknowledgments
The authors gratefully acknowledge stimulating discussions with Mircea Trif, Scott Bender, and So Takei. This work was supported in part by the NSF under Grant No. DMR-0840965 and FAME (an SRC STARnet center sponsored by MARCO and DARPA).

\appendix

\begin{widetext}

\section{Derivation of the generalized current}
\label{dgc}
In this appendix, we detail the derivation of the formula for nonequilibrium spin current through a quantum dot attached to noncollinear ferromagnetic leads, Eq.~\eqref{neq_curr}. We begin with a Hamiltonian of the form $H=H_{L}+H_{D}+H_{T}$, which describes noninteracting collinear leads ($H_L$) coupled to a quantum dot ($H_D$) by tunneling  ($H_T$).  Recall that these are defined by
\begin{equation}
H_{L}=\sum_{\sigma k \gamma}\varepsilon_{\sigma k\gamma} c_{\sigma k  \gamma}^{\dagger}c_{\sigma k  \gamma}\,,\,\,\,H_{T}=\sum_{\sigma k\gamma}V_{\sigma k \gamma}d_\sigma^\dagger c_{\sigma k\gamma}+{\rm H.c.}\,,
\end{equation}
and $H_D$ left general. The creation and annihilation operators in the basis of lead $\gamma$ canted at angle $\theta_\gamma$
\begin{align}
d_{\uparrow k \gamma}\rightarrow d^{\prime\dagger}_{\uparrow k\gamma}&=\cos\frac{\theta}{2}d^\dagger_\uparrow+\sin\frac{\theta}{2}d^\dagger_\downarrow\,,\nonumber\\
d_{\downarrow k \gamma}\rightarrow d^{\prime\dagger}_{\downarrow k\gamma}&=-\sin\frac{\theta}{2}d^\dagger_\uparrow+\cos\frac{\theta}{2}d^\dagger_\downarrow\,.
\end{align}

Both the lead and dot Hamiltonians are invariant under this transformation, while the tunneling term in the Hamiltonian becomes
\eq{H_{T}=\sum_{\sigma\sigma\prime k\gamma} d_{\sigma}^\dagger V_{\sigma\sigma^\prime k \gamma} c_{\sigma^\prime k \gamma}+{\rm H.c.}\,,} 
where
\begin{equation}
\hat{V}_{k\gamma}=V_{\sigma\sigma\prime k \gamma}\equiv
\left(
\begin{array}{cc}
V_{\uparrow k\gamma}\cos\frac{\theta_\gamma}{2} & -V_{\downarrow k\gamma}\sin\frac{\theta_\gamma}{2} \\ 
V_{\uparrow k\gamma}\sin\frac{\theta_\gamma}{2} & V_{\downarrow k\gamma}\cos\frac{\theta_\gamma}{2}
\end{array}
\right) \,.
\end{equation}
Closely following the approach of Ref.~\onlinecite{meirPRL92}, we begin by taking the time derivative of $d^\dagger_\mu d_\nu$ 
\begin{align}
-i\hbar\frac{d}{d t}(d^\dagger_\mu d_\nu)&=[H,d^\dagger_\mu d_\nu]=\frac{i}{\hbar}[H_D+H_T,d^\dagger_\mu d_\nu]=[H_D+H_T,d^\dagger_\mu]d_\nu+d^\dagger_\mu[H_D+H_T,d_\nu]\nonumber\\
&=(\epsilon_\nu-\epsilon_\mu)d_\mu^\dagger d_\nu+\sum_{\alpha k\gamma}\left(V^{\dagger}_{\alpha \mu k\gamma}c^\dagger_{\alpha k\gamma} d_\nu - V_{\nu \alpha k\gamma}d^\dagger_\mu c_{\alpha k\gamma}\right)\,.
\label{dens_dot}
\end{align}
We equate the expectation value of Eq.~(\ref{dens_dot}), without  magnetic precision, to the current
\begin{equation}
J^\gamma_{\mu\nu} = \frac{i}{\hbar}\sum_{\alpha k}\left(V^{\dagger}_{\alpha\mu k\gamma}\av{c^\dagger_{\alpha k\gamma} d_\nu} - V_{\nu\alpha k\gamma}\av{d^\dagger_\mu c_{\alpha k\gamma}}\right)=\frac{1}{\hbar} \int\frac{d\omega}{2\pi} \sum_{\alpha k} \left[V^{\dagger}_{\alpha\mu k\gamma}G_{\nu, k\alpha\gamma}^< (\omega) - V_{\nu\alpha k\gamma}G_{\alpha k\gamma,\mu}^< (\omega)\right]\,,
\end{equation}
where we have defined $G^<_{\nu,\alpha k\gamma}(t)= i\langle c^\dagger_{\alpha k\gamma} d_\nu\rangle(t)$ and $G^<_{\alpha k\gamma,\mu}(t) = i\langle d^\dagger_\nu c_{\alpha k\gamma}\rangle (t)$ which are lesser Green's functions in the Keldysh formalism. Taking advantage of the noninteracting electrons in the leads, we use their equations of motion and Langreth's theorem\cite{haugBK08} to separate the lead and dot Green's functions
\begin{equation}
G^<_{\nu,\alpha k\gamma}=-\sum_{\lambda} V_{\lambda\alpha k\gamma}\left( g^{\bar t}_{\alpha k\gamma} G^<_{\nu\lambda} - g^<_{\alpha k\gamma} G^{t}_{\nu\lambda} \right)\,,\,\,\,G^<_{\alpha k\gamma,\mu}=\sum_{\lambda} V^\dagger_{\alpha\lambda k\gamma}\left( g^t_{\alpha k\gamma} G^<_{\lambda\mu} - g^<_{\alpha k\gamma} G^{\bar t}_{\lambda\mu} \right)\,,
\end{equation}
where we have introduced the (anti-)time-ordered Green's functions on lead $\gamma$ ($g^{\bar t}_{\alpha k\gamma}$) $g^t_{\alpha k\gamma}$ of momentum $k$ and spin $\alpha$ and on the dot ($G^{\bar t}_{\mu\nu}$) $G^t_{\mu\nu}$ from spin $\mu$ to spin $\nu$. Plugging this back into our equation for current we obtain
\begin{equation}
J^\gamma_{\mu\nu} = -\frac{1}{\hbar} \sum_{\alpha k}\int\frac{d\omega}{2\pi}\left[  V^{\dagger}_{\alpha\mu k\gamma}V_{\lambda\alpha k\gamma}\left( g^{\bar t}_{\alpha k\gamma} G^<_{\nu\lambda} - g^<_{\alpha k\gamma} G^{t}_{\nu\lambda} \right)+V_{\nu\alpha k\gamma}V^\dagger_{\alpha\lambda k\gamma}\left( g^t_{\alpha k\gamma} G^<_{\lambda\mu} - g^<_{\alpha k\gamma} G^{\bar t}_{\lambda\mu}\right)\right]\,.
\end{equation}
Using identities between Green's functions \cite{mahanBK93} we rewrite the expression for the current, eliminating the time-ordered Green's functions in favor of retarded and advanced Green's functions
\begin{equation}
J^\gamma_{\mu\nu} = -\frac{1}{\hbar} \sum_{\alpha k}\int\frac{d\omega}{2\pi}\left[-V^{\dagger}_{\alpha\mu k\gamma}V_{\lambda\alpha k\gamma}\left( g^{A}_{\alpha k\gamma} G^<_{\nu\lambda} + g^<_{\alpha k\gamma} G^{R}_{\nu\lambda} \right) + V_{\nu\alpha k\gamma}V^\dagger_{\alpha\lambda k\gamma}\left( g^{R}_{\alpha k\gamma} G^<_{\lambda\mu} + g^<_{\alpha k\gamma} G^{A}_{\lambda\mu} \right)\right]\,,
\end{equation}
where $G^R_{ij}=-i\Theta(t-t')\langle \{d_i(t),d_j^\dagger(t')\}\rangle$ and $G^A_{ij}=(G^R)^*_{ji}$. After some tedious but straightforward manipulations one may show that 
\begin{equation}
\sum_{\mu\nu\alpha\lambda}\left[-V^{\dagger}_{\alpha\mu k\gamma}V_{\lambda\alpha k\gamma}\left( g^{A}_\alpha G^<_{\nu\lambda} + g^<_{\alpha\gamma} G^{R}_{\nu\lambda} \right)\right]^*\sigma_{\mu\nu}^*=\sum_{\mu\nu\alpha\lambda}\left[V_{\nu\alpha k\gamma}V^\dagger_{\alpha\lambda k\gamma}\left( g^{R}_{\alpha\gamma} G^<_{\lambda\mu} + g^<_{\alpha\gamma} G^{A}_{\lambda\mu}\right)\right]\sigma_{\mu\nu}\,,
\end{equation}
and therefore the spin current is real. Note that the free electron Green's functions uncoupled to the dot are
\begin{equation}
g^<_{\sigma k \gamma}=2\pi i \delta (\omega - \varepsilon_{\sigma k \gamma}) f_{\sigma\gamma}(\omega)\,,\,\,\,g^R_{\sigma k \gamma}=\frac{1}{\omega-\varepsilon_{\sigma k \gamma}+i\eta}=-i\pi \delta(\omega-\varepsilon_{\sigma k \gamma})+\mP\frac{1}{\omega-\varepsilon_{\sigma k \gamma}}\,,
\end{equation}
where $\eta\rightarrow0^+$ and $\mP$ denotes the Cauchy principal value. We define 
\begin{align}
\hat{W}^R_{\gamma}&=-\frac{i}{2}\left(
\begin{array}{cc}
\Gamma_{\uparrow\gamma}\cos^2\frac{\theta_\gamma}{2}+\Gamma_{\downarrow\gamma}\sin^2\frac{\theta_\gamma}{2} & \frac{1}{2}(\Gamma_{\uparrow\gamma}-\Gamma_{\downarrow\gamma})\sin\theta_\gamma \\ 
\frac{1}{2}(\Gamma_{\uparrow\gamma}-\Gamma_{\downarrow\gamma})\sin\theta_\gamma & \Gamma_{\downarrow\gamma}\cos^2\frac{\theta_\gamma}{2}+\Gamma_{\uparrow\gamma}\sin^2\frac{\theta_\gamma}{2}
\end{array}
\right)\,,
\end{align}
where $\Gamma_{\sigma\gamma}=2\pi\sum_k \left|V_{\sigma k \gamma}\right|^2 \delta(\omega-\varepsilon_{\sigma k \gamma})=2\pi\int \left|V_{\sigma k \gamma}\right|^2 \rho_{\sigma\gamma} \delta(\omega-\varepsilon_{\sigma k \gamma})= 2\pi\left|V_{\sigma\gamma}\right|^2 \rho_{\sigma\gamma}$ so that $\sum_{\alpha k}V_{\lambda\alpha k\gamma}V^\dagger_{\alpha\mu k\gamma}\mbox{Im}\,g^R_{\alpha\gamma}=-\sum_{\alpha k}V_{\lambda\alpha k\gamma}V^\dagger_{\alpha\mu k\gamma}\mbox{Im}\,g^A_{\alpha\gamma}=W^R_{\lambda\mu\gamma}$. Likewise defining $W^<_{\lambda\mu\gamma}=\sum_{\alpha k}V_{\lambda\alpha k\gamma}V^\dagger_{\alpha\mu k\gamma}g^<_{\alpha\gamma}$, we find
\begin{align}
\hat{W}^<_{\gamma}&=i\left(
\begin{array}{cc}
f_{\uparrow\gamma}(\omega)\Gamma_{\uparrow\gamma}\cos^2\frac{\theta_\gamma}{2}+f_{\downarrow\gamma}(\omega)\Gamma_{\downarrow\gamma}\sin^2\frac{\theta_\gamma}{2} & \frac{1}{2}[f_{\uparrow\gamma}(\omega)\Gamma_{\uparrow\gamma}-f_{\downarrow\gamma}(\omega)\Gamma_{\downarrow\gamma}]\sin\theta_\gamma \\ 
\frac{1}{2}\left[f_{\uparrow\gamma}(\omega)\Gamma_{\uparrow\gamma}-f_{\downarrow\gamma}(\omega)\Gamma_{\downarrow\gamma})\right]\sin\theta_\gamma & f_{\downarrow\gamma}(\omega)\Gamma_{\downarrow\gamma}\cos^2\frac{\theta_\gamma}{2}+f_{\uparrow\gamma}(\omega)\Gamma_{\uparrow\gamma}\sin^2\frac{\theta_\gamma}{2}
\end{array}
\right)\,,
\end{align}
which reduces to $W^<_{\nu\lambda\gamma}=-2f_\gamma(\omega) W^R_{\nu\lambda\gamma}$ in the absence of spin biasing. $f_{\sigma\gamma}(\omega)$ is the Fermi-Dirac distribution function in lead $\gamma$ with spin $\sigma$. We take $\Gamma_{\sigma\gamma}$ to be energy independent inside the electron band of the leads and zero otherwise. The principal part can be evaluated as
\begin{equation}
\begin{aligned}
&\sum_{\alpha k}\int d\omega V_{\lambda\alpha k\gamma}V^{\dagger}_{\alpha\mu k\gamma}G_{\nu\lambda}^<\mP\frac{1}{\omega-\varepsilon_{\sigma k \gamma}}=\sum_{\alpha k}\mP\int d\omega d\omega'V_{\lambda\alpha k\gamma}V^{\dagger}_{\alpha\mu k\gamma}G_{\nu\lambda}^<(\omega)\frac{\delta(\omega'-\varepsilon_{\sigma k \gamma})}{\omega-\omega'}\\
&\hspace{1cm}=\sum_{\alpha k}\mP\int d\omega d\omega'V_{\lambda\alpha k\gamma}V^{\dagger}_{\alpha\mu k\gamma}G_{\nu\lambda}^<(\omega')\frac{\delta(\omega'-\varepsilon_{\sigma k \gamma})}{\omega'-\omega}=-W^R_{\lambda\mu\gamma}\int d\omega\mP\int\frac{ d\omega'}{i\pi}\frac{G_{\nu\lambda}^<(\omega')}{\omega'-\omega}\,.
\end{aligned}
\end{equation}
The spin current can be written in the compact form
\begin{equation}
\begin{aligned}
\textbf{J}_\gamma &= \textbf{S}_{\mu\nu} J^\gamma_{\mu\nu} = \frac{\hbar}{2} \boldsymbol{\sigma}_{\mu\nu} J^\gamma_{\mu\nu}=\mbox{Re} \sum_{\mu\nu\lambda} \int \frac{ d\omega}{2\pi}\left[ -W^{R}_{\lambda\mu\gamma} \boldsymbol{\sigma}_{\mu\nu} G^<_{\nu\lambda} + W^{<}_{\lambda\mu\gamma}\boldsymbol{\sigma}_{\mu\nu} G^{R}_{\nu\lambda}+iW^R_{\lambda\mu\gamma}\boldsymbol{\sigma}_{\mu\nu}\mP\int\frac{ d\omega'}{\pi}\frac{G_{\nu\lambda}^<}{\omega'-\omega}\right]\\
&= \mbox{Re}\int \frac{ d\omega}{2\pi} \textnormal{Tr}\left[\hat W^{<}_\gamma\hat{\boldsymbol{\sigma}} \hat G^{R}-\hat W^{R}_\gamma \hat{\boldsymbol{\sigma}} \left(\hat G^<-\frac{i}{\pi}\mP\int d\omega' \frac{\hat{G}^<(\omega')}{\omega'-\omega}\right)\right]\,,
\label{app_Js}
\end{aligned}
\end{equation}
where $\hat{W}_\gamma^R\equiv W^R_{\mu\nu\gamma}$, $\hat{G^<}\equiv G^<_{\mu\nu}$, $\hat{W}_\gamma^<\equiv W^<_{\mu\nu\gamma}$, and $\hat{G^R}\equiv G^R_{\mu\nu}$. Using the cyclic properties of the trace and that $(\hat{G}^<)^\dagger=-\hat{G}^<$ and $(i\hat{W}^R_\gamma)^\dagger=i\hat{W}^R_\gamma$ we recover the known result for charge current \cite{meirPRL92}
\eq{
J_\gamma=e\delta_{\mu\nu}J^\gamma_{\mu\nu}=\frac{2e}{\hbar}\mbox{Re}\int \frac{ d\omega}{2\pi} \textnormal{Tr}\left(\hat W^{<}_\gamma \hat G^{R}-\hat W^{R}_\gamma \hat G^<\right)\,.
}
These general expressions for charge and spin current under voltage, temperature, or spin bias is the principal result of our work. A similar expression was derived in Ref.~\onlinecite{zhangJPCM09} for spin-independent biasing.

\section{Lesser Green's function}
\label{lgf}

In order to determine the on-site lesser Green's function, $\hat G^<=\hat G^R\hat{\Sigma}^<\hat G^A$, out of equilibrium we use a generalized\cite{swirkowiczPRB06} Ng's ansatz:\cite{ngPRL96}
\eq{\hat\Sigma^< = -\frac{1}{2}\left(\hat W^<_l + \hat W^<_r\right)\left(\hat W^R_l + \hat W^R_r\right)^{-1}\left[(\hat G^R)^{-1}-(\hat G^A)^{-1}\right]\,.}
The advantage of this approach is that we recover the exact expression for $\hat{G}^<$ when either no bias is applied or there is no interaction on the dot.

Although this does not simplify the general form of the current, considering only voltage and temperature biasing, from which it follows $\hat{W}^<=-2f_\gamma(\omega) \hat{W}^R$, the lesser Green's function simplifies to
\eq{\hat{G}^<=\hat{G}^R\left(f_l\hat W^R_l + f_r\hat W^R_r\right)\left(\hat W^R_l + \hat W^R_r\right)^{-1}\left[(\hat G^R)^{-1}-(\hat G^A)^{-1}\right]\hat{G}^A\,}
and the spin current takes the form
\begin{equation}
\begin{aligned}
\textbf{J}_\gamma=&\textnormal{Re}\int \frac{ d\omega}{2\pi} \textnormal{Tr}\bigg[\hat W^{A}_\gamma \hat\sigma^i \hat G^R\left(f_l\hat W^R_l + f_r\hat W^R_r\right)\left(\hat W^R_l + \hat W^R_r\right)^{-1}\left[(\hat G^R)^{-1}-(\hat G^A)^{-1}\right]\hat{G}^A\\
&+ f_\gamma\hat W^{R}_\gamma\hat{\boldsymbol{\sigma}}\left(\hat G^{R}-\hat G^A\right)+f_\gamma \hat W^R_\gamma\hat{\boldsymbol{\sigma}}\left(\hat G^R+\hat G^A\right)-\hat W^{R}_\gamma \hat{\boldsymbol{\sigma}}\frac{i}{\pi}\mP\int d\omega' \frac{\hat{G}^<(\omega')}{\omega'-\omega}\bigg]\,.
\end{aligned}
\end{equation}
Using $\hat G^R-\hat G^A=\hat G^R[(\hat G^A)^{-1}-(\hat G^R)^{-1}]\hat G^A$ and inserting the identity $1=(\hat W^R_l + \hat W^R_r)(\hat W^R_l + \hat W^R_r)^{-1}$, we have
\begin{equation}
\begin{aligned}
\textbf{J}_\gamma=&-\textnormal{Re}\int \frac{ d\omega}{2\pi}\left(f_{\bar\gamma}-f_\gamma\right) \textnormal{Tr}\left[\hat W^{R}_\gamma \hat{\boldsymbol{\sigma}} \hat{G}^R\hat W^R_{\bar\gamma}\left(\hat W^R_l + \hat W^R_r\right)^{-1}\left[(\hat G^R)^{-1}-(\hat G^A)^{-1}\right]\hat{G}^A\right]\\
&+\textnormal{Re}\int\frac{ d\omega}{2\pi} \textnormal{Tr}\left[f_\gamma\hat W_\gamma^{R}\hat{\boldsymbol{\sigma}} \left(\hat G^R + \hat G^A\right)-\hat W^{R}_\gamma \hat{\boldsymbol{\sigma}}\frac{i}{\pi}\mP\int d\omega' \frac{\hat{G}^<(\omega')}{\omega'-\omega}\right]\,.
\end{aligned}
\end{equation}
wherein the spin current is clearly separated between equilibrium and nonequilibrium contributions.

\section{Anderson Green's function}
\label{agf}

We calculate the time-ordered Green's function of the dot
\eq{G^t_{ij}(t-t')=\left\langle T\left\{d_i(t)d_j^\dagger(t')\right\}\right\rangle\,,}
where $T$ denotes time ordering of the operators and $\av{\cdots}$ the thermal average of $\cdots$. Using the fact that
\eq{[d_i,H]=\epsilon_i d_i  + Ud_i n_{\bar{i}}+ \sum_{\sigma k\gamma}V_{i\sigma k \gamma}c_{\sigma k\gamma}\,,}
where we have introduced $n_i=d^\dagger_i d_i$ and $\bar{i}=-i$, we follow the equation-of-motion (EoM) method\cite{haugBK08} of determining the Green's function by taking the time derivative
\eq{\begin{aligned}
i\frac{d}{dt} G^t_{ij}(t-t')&=\delta(t-t')\delta_{ij}+\left\langle T\left\{[d_i,H](t),d_j^\dagger(t')\right\}\right\rangle \\
&=\delta(t-t')\delta_{ij}+\left\langle T\left\{[\epsilon_i d_i  + Ud_i n_{\bar{i}}+ \sum_{\sigma k\gamma}V_{i\sigma k \gamma}c_{\sigma k\gamma}](t),d_j^\dagger(t')\right\}\right\rangle \\
&=\delta(t-t')\delta_{ij} + \epsilon_i G^t_{ij}(t-t') + U \Gamma_{i}  + \sum_{\sigma k\gamma }V_{i\sigma k\gamma}G^t_{\sigma k\gamma, j}(t-t')\,,
\end{aligned}}
where $G^t_{\sigma k\gamma, j}(t-t')=\langle T \{c_{\sigma k \gamma}(t),d_j^\dagger(t')\}\rangle$ and $\Gamma_{i}=\langle\langle d_i n_{\bar{i}},d_j^\dagger\rangle\rangle$. We adopt the notation that $\langle\langle A,B\rangle\rangle=\left\langle T\{A(t),B(t')\}\right\rangle$.  We further differentiate $G^t_{\sigma k\gamma, j}(t-t')$
\eq{i\frac{d}{dt} G^t_{\sigma k\gamma, j}(t-t') = \varepsilon_{\sigma k \gamma} G^t_{\sigma k\gamma, j}(t-t') + \sum_{i}V^\dagger_{\sigma i k \gamma}G^t_{ij}(t-t')\,,}
or, after Fourier transforming,
\eq{(\omega-\varepsilon_{\sigma k\gamma})G^t_{\sigma k\gamma, j}(\omega)=\sum_{i}V^\dagger_{\sigma i k\gamma}G^t_{i j}(\omega)\,,}
where we have used the commutation relation $[c_{\sigma k \gamma},H]=\varepsilon_{\sigma k \gamma} c_{\sigma k\gamma} + V^\dagger_{\sigma\sigma' k\gamma}d_{\sigma'}$. Solving for the Green's function
\eq{G^t_{\sigma k \gamma, j}(\omega)=\frac{\sum_{i}V^\dagger_{\sigma i k\gamma}G^t_{i j}(\omega)}{\omega-\varepsilon_{\sigma k \gamma}}\,.}
The equation of motion for $\Gamma_{i}$ is
\eq{i\frac{d}{dt}\Gamma_i=\delta(t-t')\left\langle\left\{d_i n_{\bar{i}},d^\dagger_j\right\}\right\rangle +\left\langle\left\langle[d_i n_{\bar{i}},H],d_j^\dagger\right\rangle\right\rangle\,.}
Taking these (anti-)commutation relations gives
\eq{
\left\{d_i n_{\bar{i}},d^\dagger_j\right\}=\delta_{ij}n_{\bar{i}}-\delta_{\bar{i}j}d^\dagger_{\bar{i}}d_i\,,\,\,\,\left[d_i n_{\bar{i}},H\right]=(\epsilon_i + U) d_i n_{\bar{i}} + \sum_{\sigma k \gamma}\left(d_i d^\dagger_{\bar{i}}V_{\bar{i}\sigma}c_{\sigma k\gamma} - d_i V^\dagger_{\sigma\bar{i}}c^\dagger_{\sigma k \gamma}d_{\bar{i}}+ V_{i\sigma k\gamma}c_{\sigma k\gamma}n_{\bar{i}}\right)\,.
}
In order to time differentiate these new operators we will again need to commute them with the Hamiltonian:
\eq{\begin{aligned}
\left[d_i d^\dagger_{\bar{i}}c_{\sigma k\gamma},H\right]=&(\epsilon_i - \epsilon_{\bar{i}} + \varepsilon_{\sigma k\gamma})d_i d^\dagger_{\bar{i}}c_{\sigma k\gamma} + d_i d^\dagger_{\bar{i}}d_\mu V^\dagger_{\sigma\mu k\gamma} + \sum_{\mu k'\gamma'} \left(V_{i\mu k'\gamma'}c_{\mu k'\gamma'} d^\dagger_{\bar{i}}c_{\sigma k\gamma} - d_i V^\dagger_{\mu\bar{i} k'\gamma'}c^\dagger_{\mu k'\gamma'}c_{\sigma k\gamma}\right)\,,\\
\left[d_i c^\dagger_{\sigma k\gamma}d_{\bar{i}},H\right]=&(\epsilon_i + \epsilon_{\bar{i}} - \varepsilon_{\sigma k\gamma} + U)d_i c^\dagger_{\sigma k\gamma}d_{\bar{i}} - d_i d_{\bar{i}} V_{\mu\sigma  k\gamma}d^\dagger_\mu  \sum_{\mu k'\gamma'} \left(d_i V_{\bar{i}\mu}c_{\mu k'\gamma'} c^\dagger_{\sigma k\gamma} + V_{i\mu  k'\gamma'}c_{\mu k'\gamma'} d_{\bar{i}}c^\dagger_{\sigma k\gamma}\right)\,,\\
\left[c_{\sigma k\gamma}n_{\bar{i}},H\right]=&\varepsilon_{\sigma k\gamma}c_{\sigma k\gamma}n_{\bar{i}} +  n_{\bar{i}}V^\dagger_{\sigma\mu k\gamma}d_\mu + \sum_{\mu k'\gamma'}\left(d^\dagger_{\bar{i}}V_{\bar{i}\mu k'\gamma'}c_{\mu k'\gamma'} c_{\sigma k\gamma} - V^\dagger_{\mu\bar{i} k'\gamma'} c^\dagger_{\mu k'\gamma'} d_{\bar{i}} c_{\sigma k\gamma}\right)\,.
\label{commute}
\end{aligned}}
Evaluating the anticommutator of these operators with $d^\dagger_j$
\eq{
\left\{d_i d^\dagger_{\bar{i}}c_{\sigma k\gamma},d^\dagger_j\right\}=\delta_{ij}d^\dagger_{\bar{i}}c_{\sigma k\gamma}\,,\,\,\,\left\{d_i c^\dagger_{\sigma k\gamma}d_{\bar{i}},d^\dagger_j\right\}=\delta_{\bar{i}j}d_i c^\dagger_{\sigma k\gamma}+\delta_{ij}c^\dagger_{\sigma k\gamma}d_{\bar{i}}\,,\,\,\,\left\{c_{\sigma k\gamma}n_{\bar{i}},d^\dagger_j\right\}=\delta_{\bar{i}j}c_{\sigma k\gamma}d^\dagger_{\bar{i}}\,.
}
Because a factor of $U$ enters the equation of motion in the second term in Eq.~(\ref{commute}), it will not contribute to the equation of motion for the full dot Green's function when the on-site interaction is large in comparison to all other energy scales in the problem. The time differentiation of the remaining operators' respective Green's functions is
\begin{align}
(\omega - \varepsilon_{\sigma k\gamma} - \epsilon_i + \epsilon_{\bar{i}})\Gamma^{(1)}_{i\sigma k\gamma}=&-\delta_{ij}\av{d^\dagger_{\bar{i}}c_{\sigma k\gamma}}-V^\dagger_{\sigma\bar{i} k\gamma}\Gamma_{i}+\sum_{\mu k'\gamma'} \left(V_{i\mu k'\gamma'}\aav{d^\dagger_{\bar{i}} c_{\mu k'\gamma'} c_{\sigma k\gamma},d_j^\dagger} + V^\dagger_{\mu\bar{i} k'\gamma'}\aav{d_i c^\dagger_{\mu k'\gamma'}c_{\sigma k\gamma},d_j^\dagger}\right)\,,\\
(\omega-\varepsilon_{\sigma  k\gamma})\Gamma^{(2)}_{\bar i\sigma  k\gamma}=&-\delta_{\bar{i}j}\av{d^\dagger_{\bar{i}}c_{\sigma k\gamma}}+V^\dagger_{\sigma i  k\gamma} \Gamma_i + \sum_{\mu k'\gamma'} \left(V_{\bar{i}\mu k'\gamma'}\aav{d^\dagger_{\bar{i}} c_{\mu k'\gamma'} c_{\sigma k\gamma},d_j^\dagger} + V^\dagger_{\mu\bar{i} k'\gamma'} \aav{d_{\bar{i}} c^\dagger_{\mu k'\gamma'}c_{\sigma k\gamma},d_j^\dagger}\right)\,,
\end{align}
where we have made use of $(d_i)^2=0$ and defined $\Gamma^{(1)}_{i\sigma k\gamma}=\aav{d^\dagger_{\bar{i}} d_i c_{\sigma k\gamma},d_j^\dagger}$ and $\Gamma^{(2)}_{\bar i\sigma k\gamma}=\aav{n_{\bar{i}}c_{\sigma k\gamma},d_j^\dagger}$. After making the following decoupling \cite{haugBK08}
\begin{align}
\aav{ c^\dagger_{\mu k'\gamma'}c_{\sigma k'\gamma'} d_i,d_j^\dagger}&\approx \av{c^\dagger_{\mu k'\gamma'}c_{\sigma k'\gamma'}}\aav{d_i,d_j^\dagger}-\av{c^\dagger_{\mu k'\gamma'}d_i}\aav{c_{\sigma k'\gamma'},d_j^\dagger}\,,\nonumber\\
\aav{d^\dagger_{\bar{i}}c_{\mu k'\gamma'} c_{\sigma k'\gamma'},d_j^\dagger}&\approx \av{d^\dagger_{\bar{i}}c_{\mu k'\gamma'}}\aav{c_{\sigma k'\gamma'},d_j^\dagger}-\av{d^\dagger_{\bar{i}}c_{\sigma k'\gamma'}}\aav{c_{\mu k'\gamma'},d_j^\dagger}\,.
\label{decouple}
\end{align}
and evaluating $\av{c^\dagger_{\mu k' \gamma'} c_{\sigma k\gamma}}=\delta_{\mu\sigma}\delta_{kk'}\delta_{\gamma\gamma'}f_{\sigma\gamma}(\varepsilon_{\sigma k\gamma})$ and $\av{d^\dagger_i c_{\sigma k\gamma}}=0$, which ensures the onset of Kondo-like features, we obtain:\cite{meirPRL93,swirkowiczPRB06}
\eq{\begin{aligned}
(\omega- \varepsilon_{\sigma k\gamma} - \epsilon_i + \epsilon_{\bar{i}})\aav{d_i d^\dagger_{\bar{i}}c_{\sigma k\gamma},d_j^\dagger}&=V^\dagger_{\sigma\bar{i} k\gamma}\aav{d_i n_{\bar{i}},d_j^\dagger} - V^\dagger_{\sigma\bar{i} k\gamma}f_{\sigma k\gamma}\aav{d_i,d_j^\dagger}\,,\\
(\omega-\varepsilon_{\sigma k\gamma})\aav{c_{\sigma k\gamma}n_{\bar{i}},d_j^\dagger}&=V^\dagger_{\sigma i k\gamma} \aav{d_i n_{\bar{i}},d_j^\dagger} + V^\dagger_{\sigma\bar{i} k\gamma} f_{\sigma k\gamma}\aav{d_{\bar{i}},d_j^\dagger}\,.
\end{aligned}}
Plugging this back into the equation of motion for $\Gamma_i$
\eq{
\omega\Gamma_i=\delta_{ij}\av{n_{\bar{i}}} - \delta_{\bar{i}j}\av{d^\dagger_{\bar{i}}d_i} + (\epsilon_i + U)\Gamma_i + \sum_{\sigma k\gamma}\frac{V_{\bar{i}\sigma k\gamma}V^\dagger_{\sigma \bar{i}k\gamma}[\Gamma_i-f_{\sigma\gamma}(\varepsilon_{\sigma k\gamma})G^t_{ij}]}{\omega  - \varepsilon_{\sigma k\gamma} - \epsilon_i + \epsilon_{\bar{i}}}+ \sum_{\sigma k\gamma}\frac{V_{i\sigma k\gamma}[V^\dagger_{\sigma i k\gamma}\Gamma_i + f_{\sigma \gamma}(\varepsilon_{\sigma k\gamma})V^\dagger_{\sigma \bar{i}k\gamma}G^t_{\bar{i}j}]}{\omega - \varepsilon_{\sigma k\gamma}}\,.
}
Owning to the strength of the on-site interaction
\eq{U\Gamma_i = \delta_{\bar{i}j}\av{d^\dagger_{\bar{i}}d_i} - \delta_{ij}\av{n_{\bar{i}}} + \sum_{\sigma k\gamma}\frac{f_{\sigma \gamma}(\varepsilon_{\sigma k\gamma})V_{\bar{i}\sigma k\gamma}V^\dagger_{\sigma \bar{i} k\gamma}}{\omega - \varepsilon_{\sigma k\gamma} - \epsilon_i + \epsilon_{\bar{i}}}G^t_{ij} - \sum_{\sigma k\gamma}\frac{f_{\sigma \gamma}(\varepsilon_{\sigma k \gamma})V_{i\sigma k\gamma} V^\dagger_{\sigma \bar{i} k\gamma}}{\omega - \varepsilon_{\sigma k\gamma}}G^t_{\bar{i}j}\,,}
which, in turn, is used to solve the expression for the original Green's function
\eq{(\omega - \epsilon_i)G^t_{ij} = \delta_{ij} + \delta_{\bar{i}j}\av{d^\dagger_{\bar{i}}d_i} - \delta_{ij}\av{n_{\bar{i}}} + \sum_\sigma(\Sigma_{0,i\sigma} + \Sigma'_{i\sigma})G^t_{\sigma j}\,,}
where we have defined the noninteracting (tunneling) self energy $\hat{\Sigma}_0$ and the Anderson dot self energy $\hat{\Sigma}'$ as
\eq{
\Sigma_{0,ij} = \sum_{\sigma k \gamma}\frac{V_{i\sigma k\gamma}V^\dagger_{\sigma j k\gamma}}{\omega-\varepsilon_{\sigma k \gamma}}\,,\,\,\,\Sigma'_{ii} = \sum_{\sigma k \gamma}\frac{f_{\sigma\gamma}(\varepsilon_{\sigma k \gamma})V_{\bar{i}\sigma k\gamma}V^\dagger_{\sigma \bar{i} k\gamma}}{\omega - \varepsilon_{\sigma k \gamma} - \epsilon_i + \epsilon_{\bar{i}}}\,,\,\,\,\Sigma'_{i\bar{i}} = - \sum_{\sigma k \gamma}\frac{f_{\sigma\gamma}(\varepsilon_{\sigma k \gamma})V_{i\sigma k\gamma} V^\dagger_{\sigma \bar{i} k\gamma}}{\omega - \varepsilon_{\sigma k\gamma }}\,.
}
Equivalently, we may write this as
\eq{(\hat{g}_0^{-1}-\hat\Sigma_0-\hat\Sigma')\hat{G}^t=\hat{m}\,,}
where $g_{0,ij} = \delta_{ij}(\omega - \epsilon_i)^{-1}$ is the free Green's function on the dot, $\hat G^t\equiv G_{ij}^t$ and $m_{ii}=1-\av{n_{\bar{i}}}$ and $m_{i\bar{i}}=\av{d_{\bar{i}}^\dagger d_i}$, which agrees with Ref.~\onlinecite{swirkowiczPRB06}. We must self-consistently determine the occupation on the dot by numerically evaluating
\begin{equation}
\av{n_i}=\mbox{Im}\int\frac{ d\omega}{2\pi}G^<_{ii}(\omega)\,,\,\,\,\av{d_{\bar{i}}^\dagger d_i}=-i\int\frac{ d\omega}{2\pi}G^<_{i\bar{i}}(\omega)\,.
\end{equation}

To extract the retarded $\hat{G}^R$ Green's functions we analytically continue $\omega$ to the imaginary plane $\omega\rightarrow\omega+i\eta$.  The imaginary part of $\hat{\Sigma}_0$ is equal to $\hat{W}^R$ while the real part is

\begin{align}
\textrm{Re}\Sigma_{0,ij}=\sum_{\sigma k \gamma}\mathcal{P}\frac{V_{i\sigma k\gamma}V^\dagger_{\sigma j k\gamma}}{\omega-\varepsilon_{\sigma k \gamma }}=\sum_\gamma\frac{W^R_{ij\gamma}}{i\pi}\ln\left|\frac{\omega-\mathcal{W}_e}{\omega+\mathcal{W}_h}\right|\,,
\end{align}

where we have assumed $\mathcal{W}_{h/e}\gg\mu_{\sigma\gamma}$. $\hat{\Sigma}'$ has the form
\eq{\sum_{\sigma k \gamma}\frac{f_{\sigma k\gamma}(\varepsilon_{\sigma k \gamma})V_{\bar{i}\sigma k\gamma}V^\dagger_{\sigma k\bar{i}\gamma}}{\omega + i\eta - \varepsilon_{\sigma k \gamma} - A}
=-i\pi\sum_{\sigma k \gamma}f_{\sigma\gamma}(\varepsilon_{\sigma k \gamma})V_{\bar{i}\sigma k\gamma}V^\dagger_{\sigma \bar{i} k\gamma}\left[\delta(\omega-\varepsilon_{\sigma k \gamma}-A)+\frac{1}{i\pi}\mP\frac{1}{\omega - \varepsilon_{\sigma k \gamma} - A}\right]\,.}
Defining the functions $\mG_{\sigma\gamma}(\omega)=(1/i\pi)\mP\int d\omega'f_{\sigma\gamma}(\omega')/(\omega-\omega')$, we obtain 

\begin{align}
\Sigma'_{ii}&=\frac{i}{2}\sum_\gamma\left\{\left[f_{i\gamma}(\omega-\epsilon_i+\epsilon_{\bar{i}})+\mG_{i\gamma}(\omega-\epsilon_i+\epsilon_{\bar{i}})\right]\Gamma_{\bar{i}\gamma}\cos^2\frac{\theta_\gamma}{2}+\left[f_{\bar{i}\gamma}(\omega-\epsilon_{i}+\epsilon_{\bar{i}})+\mG_{\bar{i}\gamma}(\omega-\epsilon_{i}+\epsilon_{\bar{i}})\right]\Gamma_{i\gamma}\sin^2\frac{\theta_\gamma}{2}\right\}\,,\nonumber\\ 
\Sigma'_{i\bar{i}}&=-\frac{i}{4}\sum_\gamma\left\{ \left[f_{\uparrow\gamma}(\omega)+\mG_{\uparrow\gamma}(\omega)\right]\Gamma_{\uparrow\gamma}-\left[f_{\downarrow\gamma}(\omega)+\mG_{\downarrow\gamma}(\omega)\right]\Gamma_{\downarrow\gamma})\right\}\sin\theta_\gamma\,.
\end{align}
In the zero temperature limit, these may be evaluated analytically
\eq{
\mG_{\sigma\gamma}(\omega)=\mP\int^{\mu_{\sigma\gamma}}_{-\mathcal{W}_h}\frac{ d\omega'}{i\pi}\frac{1}{\omega-\omega'}=-\frac{1}{i\pi}\ln\left|\frac{\omega-\mu_{\sigma\gamma}}{\omega+\mathcal{W}_h}\right|\,,}
which immediately gives the locations of the Kondo resonances at $\mu_{\sigma\gamma}$.

\end{widetext}

\end{document}